\documentstyle[12pt,epsfig]{article}
\renewcommand{\arraystretch}{1.4}
\begin{document}
\font\mybb=msbm10 at 12pt
\def\bb#1{\hbox{\mybb#1}}
\def\Z {\bb{Z}}
\def\R {\bb{R}}
\def\I {\bb{I}}
 \def\unit{\hbox to 3.3pt{\hskip1.3pt \vrule height 7pt width .4pt \hskip.7pt
\vrule height 7.85pt width .4pt \kern-2.4pt
\hrulefill \kern-3pt
\raise 4pt\hbox{\char'40}}}
\def\II{{\unit}}
\def\cM {{\cal{M}}}
\def\half{{\textstyle {1 \over 2}}}
\def\tfrac#1#2{{\textstyle{#1\over#2}}}

\pagestyle{empty}
\rightline{UG-13/95}
\rightline{HUB-EP-95/30}
\rightline{hep-th/9512152}
\rightline{December 1995}
\vspace{1truecm}
\centerline{\bf  Type II Duality Symmetries in Six Dimensions}
\vspace{1.2truecm}
\centerline{\bf Klaus Behrndt}
\vspace{.5truecm}
\centerline{Humboldt-Universit\"at, Institut f\"ur Physik}
\centerline{Invalidenstra\ss e 110, 10115 Berlin}
\centerline{Germany}
\vspace{.3truecm}
\centerline{and}
\vspace{.3truecm}
\centerline{{\bf E.~Bergshoeff, Bert Janssen}}
\vspace{.5truecm}
\centerline{Institute for Theoretical Physics}
\centerline{Nijenborgh 4, 9747 AG Groningen}
\centerline{The Netherlands}
\vspace{1.2truecm}
\centerline{ABSTRACT}
\vspace{.5truecm}
We discuss the different discrete
duality symmetries in six dimensions that act within
and between (i) the 10-dimensional heterotic string compactified on $T^4$,
(ii) the 10-dimensional Type IIA string compactified on $K3$ and
(iii) the 10-dimensional Type IIB string compactified on $K3$.
In particular we show that the underlying
group-theoretical structure of these
discrete duality symmetries is determined by the proper cubic group
${\cal C}/\Z_2$. Our group theoretical
interpretation leads to simple rules for constructing
the explicit form
of the different discrete Type II
duality symmetries in an arbitrary background.
The explicit duality rules we obtain
are applied to construct dual versions of the
6-dimensional chiral null model.
\vfill\eject
\pagestyle{plain}

\noindent  {\bf 1. Introduction}
\bigskip

Duality symmetries are playing an increasingly important role in
relating different types of string theories and in investigating their
non-perturbative behaviour. In many cases the
duality symmetries manifest themselves at the level of the low-energy
effective action as solution-generating transformations. Having
the application as solution-generating transformation
in mind it is important to have knowledge
of the explicit form of the duality symmetry in an arbirary curved
background. For the 10-dimensional heterotic string the
so-called $T$-duality rule has been given some time ago
by Buscher \cite{Bu1}.
This rule includes the well-known $R\rightarrow 1/R$ duality for the
special background ${\cal M}_9\times T^1$ where ${\cal M}_9$ is a
9-dimensional Minkowski space and $T^1$ a circle of radius $R$.
For the 10-dimensional Type II strings it has been shown that the
$T$ duality for the background ${\cal M}_9 \times T^1$
provides a map between the Type IIA and Type IIB
superstring \cite{Da1,Di1}. The corresponding expression in a general
curved background has
been given recently in \cite{Be1} thereby generalizing the Buscher's
rules to the Type II case. Both the Type I as well as the Type II
10-dimensional duality rules have been applied to generate new
solutions to the ten-dimensional low-energy string effective action.

In 6 dimensions an important role is played by the so-called
string/string duality \cite{Du1}. This string/string duality
relates
the weakly coupled $D=10$ heterotic string compactified on $T^4$ to the
strongly coupled Type IIA superstring compactified on $K3$ and
{\it vice-versa} \cite{Hu1,Wi1,Se1,Ha1}. Additional 6-dimensional
dualities have been considered involving the $D=10$ Type IIB superstring
compactified on $K3$ \cite{Wi1,Du2,As1}, leading to the concept
of string/string/string triality \cite{Du2}. It is the purpose of this paper
to explain the group-theoretical structure underlying the different
discrete 6-dimensional dualities.
Our analysis will lead to a simple do-it-yourself kit for constructing
the explicit form of the discrete
duality symmetries that act within and between
(i) the 10-dimensional heterotic string compactified on $T^4$,
(ii) the 10-dimensional Type IIA string compactified on $K3$ and
(iii) the 10-dimensional Type IIB string compactified on $K3$.

To explain the main idea of this paper it is instructive to
analyse the much simpler group theoretical structure
underlying the $D=10$ duality symmetries.
The main purpose of this $D=10$ analysis is not to present new results
but to examplify the more complicated situation of the duality
symmetries in 6 dimensions which is the topic of this paper.

We first consider the $D=10$ heterotic string.
The bosonic background fields are a metric $\hat g$,
an antisymmetric tensor $\hat B$ and a dilaton $\hat\phi$. The dilaton is
related to the string coupling constant $g_s$ via
$g_s = e^{\hat\phi}$. The zero-slope limit
action in the string-frame metric is given by\footnote{
We use the notation and conventions of \cite{Be1,Be2}. In
particular, fields are hatted before and unhatted after dimensional
reduction. Throughout
this paper we will use a form notation. If A is
a $p$-from and $B$ a $q$-form, then $|A|^2$ means
$A_{\mu_1\cdots \mu_p}A^{\mu_1\cdots \mu_p}$,
$AB$ means $A_{[\mu_1\cdots \mu_p} B_{\mu_{p+1}\cdots \mu_{p+q}]}$
and $dA$ means $\partial_{[\mu_1}A_{\mu_2\cdots \mu_{p+1}]}$.}

\begin{equation}
\label{heterotic}
I_{\rm heterotic} = \tfrac{1}{2}
\int d^{10}x {\sqrt {-{\hat g}}} e^{-2\hat\phi}\
\biggl\{-\hat R + 4|d\hat \phi|^2 -\tfrac{3}{4}|d\hat B|^2\biggr\}\, .
\end{equation}

We assume that the background fields have an isometry in a given, let us say,
${\underline x}$ direction. The $T$-duality rules are
easiest formulated by rewriting the 10-dimensional action
in 9 dimensions via the process of dimensional reduction \cite{Sc1,Ma1}.
Besides a 9-dimensional metric, antisymmetric tensor
and dilaton the dimensionally
reduced theory contains 2 additional vectors $A$ and $B$
and one scalar $\sigma$. The vectors $A$ and $B$ originate
from the 10-dimensional
metric and antisymmetric tensor, respectively.
The string coupling constant $g_s$ and the
compactification radius $R$ are related to the 9-dimensional
dilaton $\phi$ and the scalar $\sigma$ via

\begin{equation}
R = e^{\sigma/2}\, ,\hskip 1.5truecm g_s =
e^{\phi + \sigma/4}\, .
\end{equation}

To understand how the different discrete duality symmetries are
realized in 9 dimensions it
is enough to consider the kinetic terms of the two vector fields
$A$ and $B$ and their
coupling to $\sigma$ and $\phi$ which we write in the following
suggestive form, using the Einstein-frame metric\footnote{
Actually, the situation is a bit more complicated than described below
since also the 9-dimensional antisymmetric tensor is involved in the discrete
duality symmetries. For more details, see \cite{Be2}.}:

\begin{equation}
\label{hd=9}
I_{AB\sigma} = \tfrac{1}{2}\int d^9 x \sqrt {g} \biggl [
e^{{\vec Q}_A\cdot \vec \Phi}|dA|^2 + e^{{\vec Q}_B \cdot\vec \Phi}
|dB|^2 \biggr ]\, ,
\end{equation}
where $\vec\Phi = (\sigma,\phi)$ and

\begin{figure}[t]
\begin{center}
\mbox{\epsfig{file=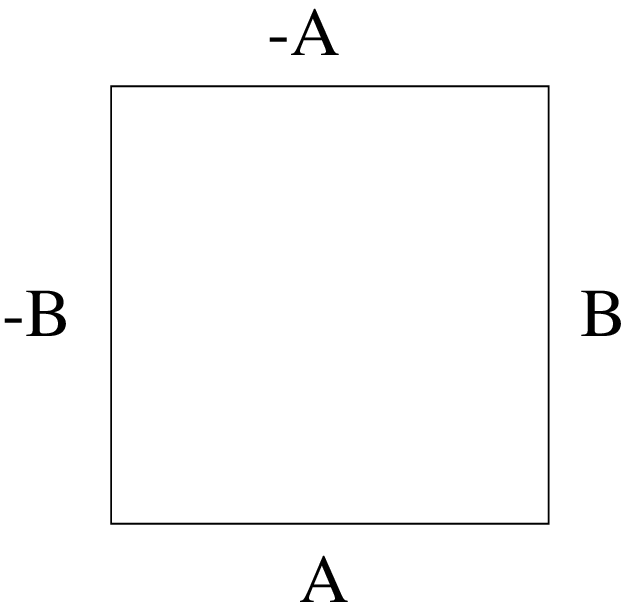, width=5cm}}
\end{center}
\noindent {\bf Fig.~1}\ \ {\it Each discrete symmetry of the square corresponds
to a symmetry acting on the 2 vectors $A$ and $B$.
The four sides of the square correspond to the pairs $(A,-A)$ and $(B,-B)$.}
\end{figure}

\begin{eqnarray}
\label{2vectors}
{\vec Q}_A &=& \bigl (1, -\tfrac{4}{7}\bigr )\, ,\nonumber\\
{\vec Q}_B &=& \bigl (-1, - \tfrac{4}{7} \bigr )\, .
\end{eqnarray}
It is not difficult to analyze the discrete duality symmetries of this action.
It turns out
that on the 2 vectors $A$ and $B$ one can realize the 8-element finite
dihedral group
$D_4$ \cite{Be2}. The
easiest way to see how this group is realized is to write a square,
like in Fig.~1, with sides $(A,-A)$ and $(B,-B)$. Every discrete symmetry
of the square corresponds to a symmetry acting on the 2 vectors. For instance,
the 2--order element of $D_4$ given by
a reflection around the diagonal from the left upper corner to the right
lower corner corresponds to the transformation $A^\prime = B, B^\prime =A$.
As another example we mention the 4--order element of $D_4$
given by a (counterclockwise) rotation over 90 degrees.
It corresponds to the symmetry $A^\prime = B, B^\prime = -A$.

To see which discrete group is realized on $\sigma$, we write
the 2 vectors $\vec Q_A$ and $\vec Q_B$ as
the corners of a 1-dimensional line--segment, like in Fig.~2.
\begin{figure}[t]
\begin{center}
\mbox{\epsfig{file=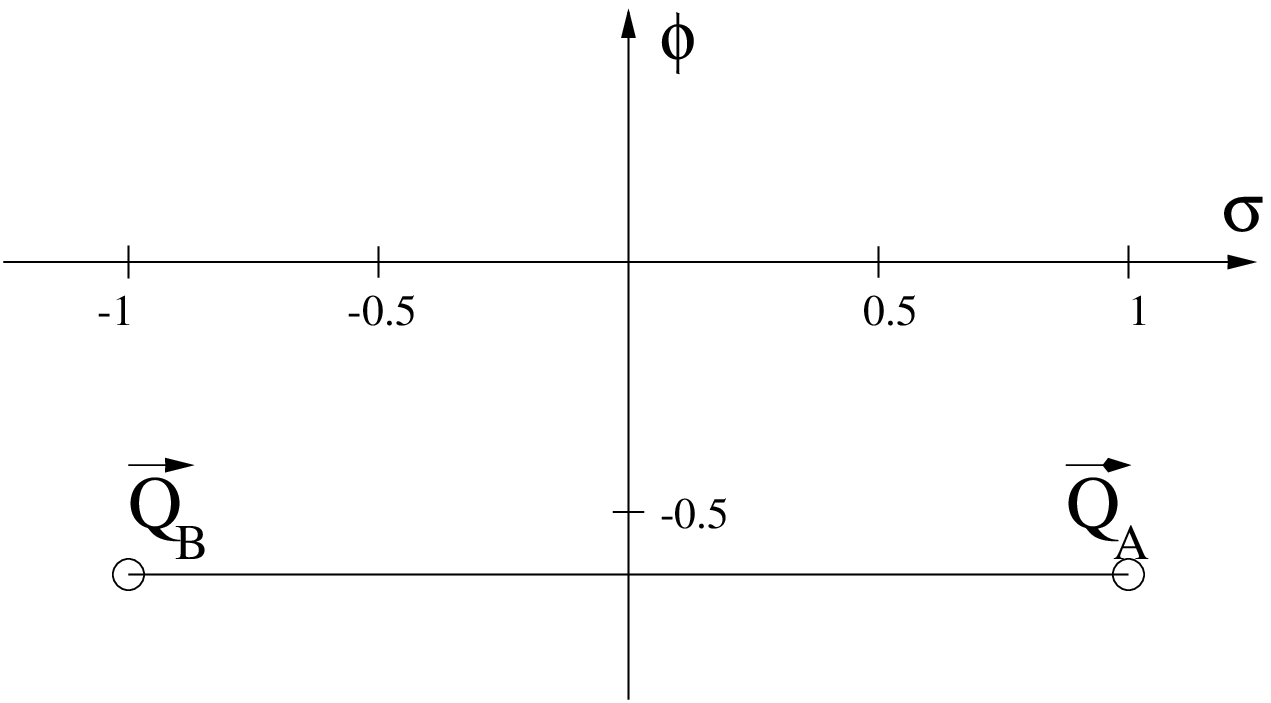, width=7cm}}
\end{center}
\noindent {\bf Fig.~2}\ \ {\it Each symmetry of the line-segment
corresponds to a symmetry acting on $\sigma, phi$. The 2 corners of the
line-element are given by the 2 vectors $\vec Q_A$ and $\vec Q_B$
defined in eq.~(\ref{2vectors}).}
\end{figure}
One thus finds that the 2-element discrete group

\begin{equation}
\Z_2 = D_4/\bigl (\Z_2\times \Z_2 \bigr )
\end{equation}
is realized on the scalar $\sigma$,
i.e.~to every 4 symmetries acting on the vectors corresponds a single
symmetry acting on the scalar. The action of $\Z_2$ on $\sigma, \phi$
is given by

\begin{eqnarray}
{\bf e}:\hskip 1.5truecm \sigma^\prime &=& \sigma\, ,\ \ \ \ \ \ \
\phi^\prime = \phi\, ,\nonumber\\
{\bf T}:\hskip 1.5truecm    \sigma^\prime &=& - \sigma\, ,\ \ \ \ \
\phi^\prime = \phi\, .
\end{eqnarray}

One may alternatively describe this $\Z_2$ duality by its action on
the compactification radius $R$ and the string coupling
constant $g_s$ as follows:

\begin{eqnarray}
{\bf e}:\hskip 1.5truecm R^\prime &=& R\, ,\ \ g_s^\prime = g_s\, ,\nonumber\\
{\bf T}: \hskip 1.5truecm   R^\prime &=& {1\over R}\, ,\ \ g_s^\prime =
{g_s\over R}\, .
\end{eqnarray}

To each element of $\Z_2$ corresponds 4 elements of $D_4$ acting
on the 2 vectors. The specific transformations of the vectors are given
by\footnote{We only give 2 elements of $D_4$. To every element
below one can associate 3 more elements  by changing (in 3 possible ways)
signs in the given transformation rules.}

\begin{eqnarray}
{\bf e}: \hskip 1.5truecm A^\prime &=& A\, , \ \ B^\prime = B\, ,\nonumber\\
{\bf T}: \hskip 1.5truecm A^\prime &=& B\, , \ \ B^\prime = A\, .
\end{eqnarray}

Combining the above $D=9$ discrete symmetries with the dimensional
reduction formulae relating the $D=9$ fields to the $D=10$ ones, it
is now straightforward to construct the $D=10$ form of the duality
rules. This leads basically to the Buscher's duality rules.
All the other discrete dualities differ from the unit transformation
or the Buscher's rules by some additional sign changes in the
10-dimensional analogues of the 9-dimensional vectors. In practice, it
often means that the sign of some charges in a given solution are
undetermined.

We now extend the above analysis to the $D=10$ Type II theories. Both
the Type IIA and Type IIB theories contain the action (\ref{heterotic})
as a common subsector. One finds that after dimensional reduction the
$D_4$ symmetry does not survive as a symmetry of the full
$D=9$ Type II action but is broken according to\footnote{
The unbroken $\Z_2$--transformation acts on the 2 vectors $A$ and $B$
as $A^\prime = -A$ and $B^\prime = -B$ \cite{Be1,Be2}.}

\begin{equation}
\label{d=9sb}
D_4 \rightarrow \Z_2\, .
\end{equation}
In particular, it turns out that the $\Z_2$ $T$-duality acting on the scalar
$\sigma$ is broken. This does {\it not} mean that there is no Type II
$T$-duality.
It just means that from the 10-dimensional point of view the $T$-duality
describes a map between {\it different} theories. In this case the
$T$-duality establishes a map between the Type IIA and Type IIB
superstring \cite{Da1,Di1}, as indicated in Fig.~3.

\begin{figure}[t]
\begin{center}
\mbox{\epsfig{file=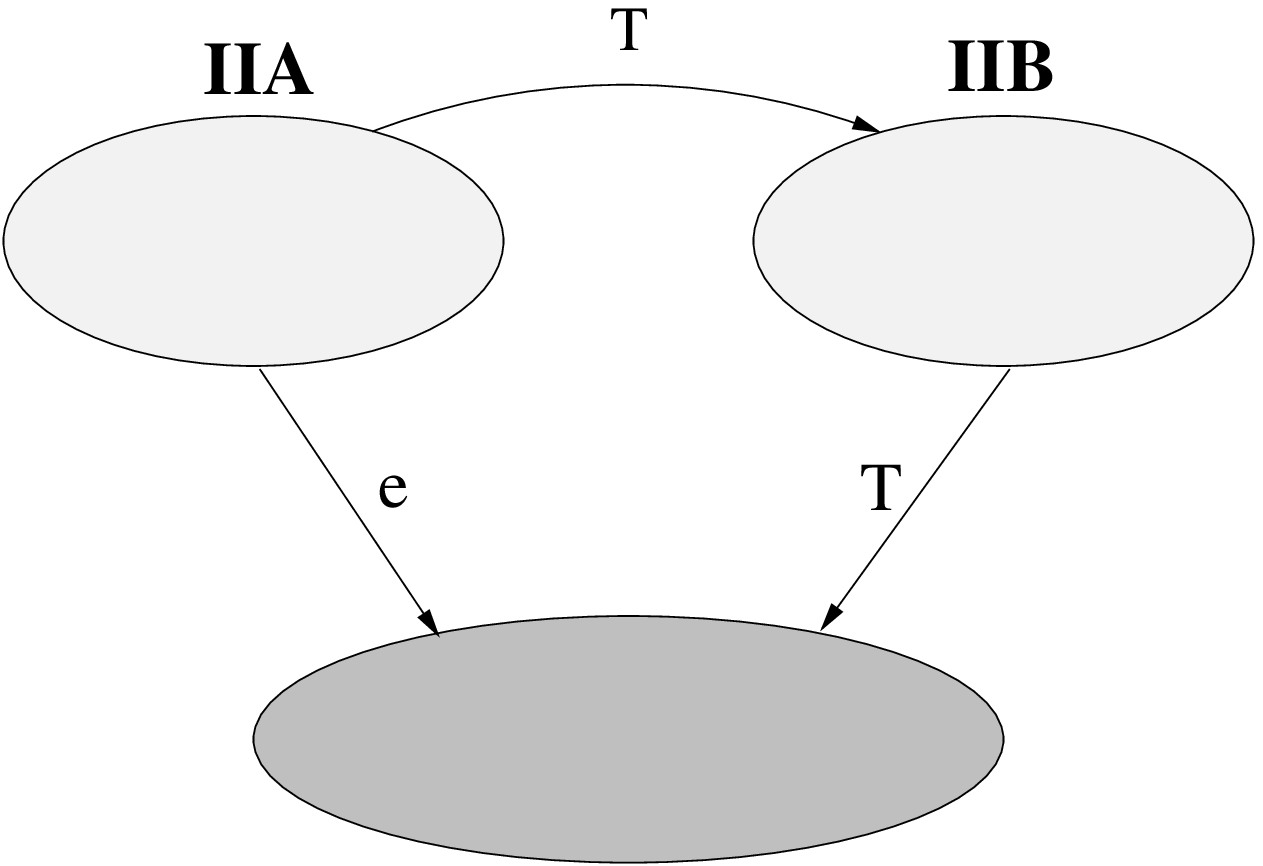, width=7cm}}
\end{center}
\noindent {\bf Fig.~3}\ \ {\it The Type II $T$--duality in ten
dimensions describes a map between the Type IIA and Type IIB
superstring. The reduction to $D=9$ of the Type IIA (Type IIB) theory is
indicated in the figure with ${\bf e}\ ({\bf T})$.}
\end{figure}

We call the reduction formula that maps the 10-dimensional
Type IIA and IIB theory onto
{\it the same} 9-dimensional theory, ${\bf e}$ and ${\bf T}$,
respectively\footnote{The corresponding inverse reduction formulae,
or oxidation formulae, are indicated by ${\bf e}^{-1}$ and ${\bf T}^{-1}$,
respectively.}
The reason for this is that, when restricted to the common subsector given
in (\ref{heterotic}), the IIB reduction formula becomes the $T$-dual
of the IIA reduction formula. For instance, for the special background,
${\cal M}_9 \times T^1$, when the IIA theory is dimensionally reduced
over a circle with radius $R_A$, the type IIB theory is dimensionally reduced
over a circle with radius $1/R_B$. Using this suggestive notation, it is
straightforward to construct the explicit form of the $D=10$ $T$-duality
rule that establishes the map from the Type IIA onto the Type IIB theory.
It is just a composition of the two reduction formulae
indicated in Figure 3. For instance, the map from IIA to IIB is obtained
by first reducing IIA with ${\bf e}$ and then oxidizing back to IIB with
${\bf T}^{-1}$. Similarly, the map from IIB to IIA is obtained by first
reducing IIB with ${\bf T}$ and then oxidizing back to IIA with ${\bf
e}^{-1}$:

\begin{eqnarray}
T (IIA \rightarrow IIB)
&=& {\bf T^{-1}} \times {\bf e} = {\bf T}\times {\bf e}= {\bf T}\, ,\nonumber
\\
T(IIB \rightarrow IIA) &=& {\bf e^{-1}}\times {\bf T} = {\bf e}\times
{\bf T} = {\bf T}\, .
\end{eqnarray}
Note that the $D=10$ Type II dualities, although not being the
symmetry of a {\it single} theory, still satisfy the $\Z_2$ group
structure in the sense that

\begin{eqnarray}
T (IIB \rightarrow IIA)\times T (IIA \rightarrow IIB)
= \I(IIA
\rightarrow IIA)\, ,\nonumber\\
T (IIA \rightarrow IIB)\times T (IIB \rightarrow IIA)
=\I(IIB \rightarrow IIB)\, .
\end{eqnarray}
This is due to our notation of the reduction formulae which is such that,
when restricted to the common subsector, each reduction formula
(and its inverse) is in 1--1 correspondence with a
specific $\Z_2$ symmetry of the $D=10$ heterotic action
(\ref{heterotic}) after dimensional reduction to $D=9$.

It is the purpose of this paper to show that the above analysis in $D=9\&
10$
can be repeated for the more complicated case in $D=5\& 6$. Basically,
apart from being technically more complicated, the only difference with the
above analysis is that different discrete duality groups are involved.
To be precise, compared with the $D=9\& 10$ situation we will
be dealing in $D=5\& 6$ with the groups indicated in Table 1.

\begin{center}
\begin{tabular}{|c|c|}
\hline
$D=9\& 10$&$D=5\& 6$\cr
\hline\hline
$D_4$ & ${\cal C}/\Z_2$\cr
$\Z_2$ & $D_3$\cr
\hline
\end{tabular}
\end{center}
\bigskip

\noindent {\bf Table 1.}\ \ {\it Discrete duality groups in
$D=9\& 10$ and $D=5\& 6$.}
\bigskip

Here ${\cal C}/\Z_2$ is the 24-element proper cubic group and $D_3$
is the 6-element dihedral group. The analogues of Figures 1,2 and 3
are given in Figures 4,5 and 6, respectively. In particular Figure 6
will provide us with simple and elegant rules for constructing the
explicit form of all discrete duality symmetries indicated in the figure.
This will turn out to be extremely useful for the purpose of constructing
new solutions corresponding to the different string effective actions involved.

The organization of this paper is as follows. In section 2 we discuss the
common sector in 6 dimensions and exhibit its discrete duality symmetries.
In section 3 we give and disscuss
the explicit form of the zero-slope limit effective
 actions
corresponding to
(i) the 10-dimensional heterotic string compactified on $T^4$,
(ii) the 10-dimensional Type IIA string compactified on $K3$ and
(iii) the 10-dimensional Type IIB string compactified on $K3$.
The dimensional reduction of these effective actions onto the {\it same}
5-dimensional Type II theory is discussed in section 4. In section 5
we show how, using the results of section 4, one may construct in
a simple way the explicit form of
the different 6-dimensional duality transformations.
Finally, in section 6 these duality rules
are applied to construct dual versions of the 6-dimensional chiral
null model.
\bigskip

\noindent  {\bf 2. The Common Sector}
\bigskip

Each of the $D=6$ string theories discussed in this paper (heterotic,
Type IIA, Type IIB) contains as a common subsector a metric $\hat g$,
an antisymmetric tensor
$\hat B$ and a dilaton $\hat \phi$. These fields define the
common sector of each string theory. In this section we will first
investigate the duality symmetries of this common sector\footnote{
The contents of this section has some overlap with \cite{Ka1}.}.
The action for the common sector
in the string-frame metric is given by

\begin{equation}
\label{commons}
I_{\rm common} = \tfrac{1}{2}\int d^6x {\sqrt {-{\hat g}}} e^{-2\hat\phi}\
\biggl\{-\hat R + 4|d\hat \phi|^2 -\tfrac{3}{4}|d\hat B|^2\biggr\}\, .
\end{equation}

The special thing about 6 dimensions is that the equations of motion
corresponding to the common sector are invariant under so-called
string/string duality transformations \cite{Du1}.
These transformations are easiest
formulated in the (6-dimensional) Einstein-frame metric

\begin{equation}
{\hat g}_E = e^{-\hat \phi} {\hat g}_S\, ,
\end{equation}
which is invariant under the string/string duality.  The action for the
common sector in the Einstein-frame metric is given by\footnote{
We will denote the string-frame and Einstein-frame metric with the
 same symbol. Whenever confusion could arise, we will explicitly
 denote with a subindex whether a metric is in Einstein- or string-frame.}:

\begin{equation}
\label{commone}
I_{\rm common} = \tfrac{1}{2}\int d^6x {\sqrt {-{\hat g}}} \
\biggl\{-\hat R - |d\hat \phi|^2 -\tfrac{3}{4}
e^{-2\hat\phi} |d\hat B|^2\biggr\}\, .
\end{equation}
The string/string duality rules in the Einstein frame are given by

\begin{equation}
{\hat \phi}^\prime = -{\hat\phi}\, ,\hskip 1.5truecm
(d\hat B)^\prime = e^{-2\hat\phi}\ {}^*(d\hat B)\, ,
\end{equation}
where

\begin{equation}
{}^*(d\hat B)_{\hat\mu\hat\nu\hat\rho} \equiv {1\over 3!{\sqrt {-{\hat g}}}}
\epsilon_{\hat\mu\hat\nu\hat\rho\hat\lambda\hat\sigma\hat\tau}
(d\hat B)^{\hat\lambda\hat\sigma\hat\tau}\, .
\end{equation}

We now discuss the reduction to five dimensions, assuming there is an
isometry in a given, let us say, $\underline x$-direction. Using both in
$D=6$ as well as $D=5$ the string--frame metric, the
6-dimensional fields are expressed in terms of the
five-dimensional ones as follows\footnote{We have used the decomposition
$\hat \mu = (\mu,{\underline x})$.}:
\begin{eqnarray}
\label{d=6d=5}
{\hat g}_{\underline{xx}} &=& - e^{-4\sigma/{\sqrt 3}}\, ,\nonumber\\
{\hat g}_{{\underline x}\mu} &=& -e^{-4\sigma/{\sqrt
 3}}A_\mu\, ,\nonumber\\
{\hat g}_{\mu\nu} &=& g_{\mu\nu} - e^{-4\sigma/{\sqrt 3}}A_\mu
 A_\nu\, ,\\
{\hat B}_{\mu\nu} &=& B_{\mu\nu} + A_{[\mu}B_{\nu]}\, ,\nonumber\\
{\hat B}_{\underline x \mu} &=& B_\mu\, ,\nonumber\\
\hat \phi &=& \phi - \tfrac{1}{{\sqrt 3}} \sigma\, .\nonumber
\end{eqnarray}
The dimensionally reduced action in the (5-dimensional)
string frame metric is given by

\begin{eqnarray}
\label{common5}
I_{common} = &&\tfrac{1}{2}\int d^5 x \sqrt {g} e^{-2\phi}\biggl [
-R + 4|d\phi|^2 -\tfrac{3}{4}|H|^2 -\tfrac{4}{3} |d\sigma|^2\\
&& +e^{-4\sigma/\sqrt 3}|dA|^2 +  e^{4\sigma/\sqrt
 3} |dB|^2 \biggr ]\, ,\nonumber
\end{eqnarray}
with
\begin{equation}
H = dB + AdB + BdA\, .
\end{equation}

We next use the fact that 5 dimensions is special in the sense that in this
dimension the antisymmetric tensor
$B$ is dual to a vector $C$ \cite{Wi1}. In terms of this
vector $C$ the action is given by\footnote{For clarity we give here
the component form
of the last (topological) term in the action:

\begin{eqnarray}
-\tfrac{1}{2}\int_{{\cal M}_6} dAdBdC &=& -\tfrac{1}{2}\int_{{\cal M}_6}
d^6 x\
\epsilon^{\hat\mu\hat\nu\hat\rho\hat\lambda\hat\sigma\hat\tau}
\partial_{\hat\mu} A_{\hat\nu}\partial_{\hat\rho}B_{\hat\lambda}
\partial_{\hat\sigma}C_{\hat\tau}\nonumber\\
&=& -\tfrac{1}{2}\int_{{\cal M}_5} d^5x\ \epsilon^{\mu\nu\rho\lambda\sigma}
A_\mu\partial_\nu B_\rho\partial_\lambda C_\sigma\, .
\end{eqnarray}
}

\begin{eqnarray}
\label{common5v}
I_{common} = &&\tfrac{1}{2}\int_{{\cal M}_5}
d^5 x \sqrt {g} e^{-2\phi}\ \biggl [
-R + 4|d\phi|^2  -\tfrac{4}{3} |d\sigma|^2\nonumber\\
&& + e^{-4\sigma/\sqrt
 3}|dA|^2 +  e^{4\sigma/\sqrt 3} |dB|^2
+  e^{4\phi}|dC|^2  \biggr ]\nonumber\\
&&-\tfrac{1}{2}\int_{{\cal M}_6} dAdBdC\, ,
\end{eqnarray}
where ${\cal M}_6$ is a 6 manifold with boundary ${\cal M}_5$.

To study the symmetries of the dimensionally reduced action it is
convenient to use the (5-dimensional) Einstein frame metric

\begin{equation}
g_E = e^{-4\phi/3} g_S\, ,
\end{equation}
in terms of which the action becomes

\begin{eqnarray}
\label{common5s}
I_{common} = &&\tfrac{1}{2}\int_{{\cal M}_5}
d^5 x \sqrt {g} \biggl [
-R - \tfrac{4}{3} |d\phi|^2  -\tfrac{4}{3} |d\sigma|^2\nonumber\\
&& +\ e^{-4\vec Q_A \cdot \vec \Phi/3}|dA|^2 +  e^{
-4\vec Q_B\cdot \vec \Phi/3} |dB|^2
+ e^{-4\vec Q_C\cdot \vec\Phi/3}|dC|^2  \biggr ]\nonumber\\
&&-\tfrac{1}{2}\int_{{\cal M}_6} dAdBdC\, ,
\end{eqnarray}
where $\vec\Phi = \bigl (\sigma, \phi \bigr)$ and

\begin{eqnarray}
\label{vector}
\vec Q_A &=&  \bigl (\sqrt 3, 1\bigr )\, ,\nonumber\\
\vec Q_B &=&  \bigl (-\sqrt 3, 1\bigr )\, ,\\
\vec Q_C &=&  \bigl ( 0, -2\bigr )\, .\nonumber
\end{eqnarray}

Given the above form of the dimensionally reduced action it is not
difficult to analyse its discrete duality symmetries. It turns out
that on the 3 vectors one can realize the 24-element finite group
${\cal C}/\Z_2$ where ${\cal C}$ is the so-called cubic group. The
easiest way to see how this group is realized is to write a cube,
like in Fig.~4, with faces $(A,-A), (B,-B)$ and $(C,-C)$.

\begin{figure}[t]
\begin{center}
\mbox{\epsfig{file=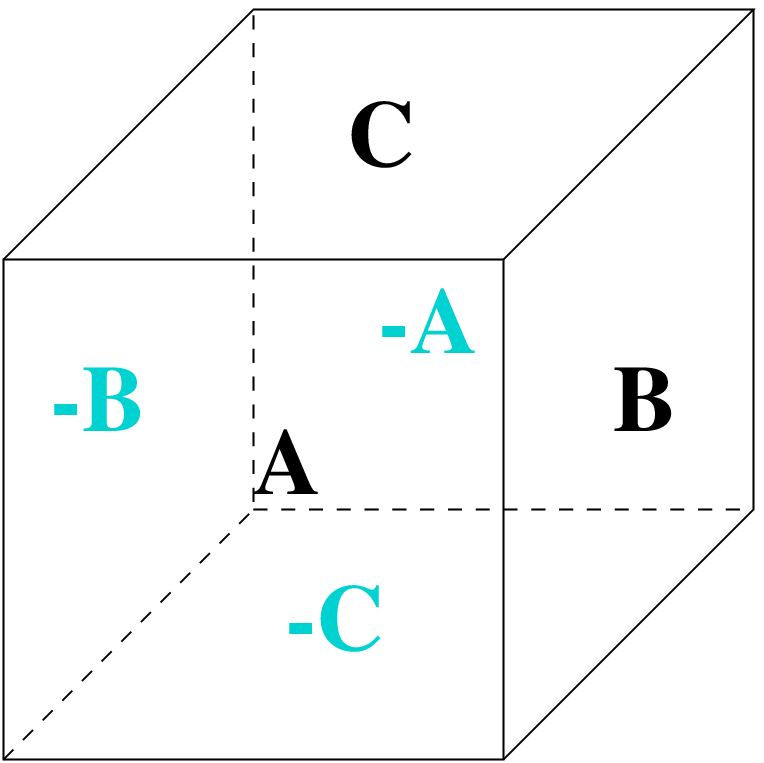, width=5cm}}
\end{center}
\noindent {\bf Fig.~4}\ \ {\it Each proper discrete
symmetry of the cube corresponds
to a symmetry acting on the 3 vectors. The six faces of the cube
correspond to the pairs $(A,-A), (B,-B)$ and $(C,-C)$.}
\end{figure}

The reason that we
only consider the 24 proper symmetries and not the full 48-element
cubic group is that only the proper elements leave the last
(topological) term in the action (\ref{common5s}) invariant.
The proper cubic group has elements of order 2 and 3\footnote{
The full cubic group also has elements of order 4. An example of
such a 4-order element is
a rotation of 90 degrees with axis the line going from the center
of the lower face to the center of the upper face of the cube.}.
An example of a 2-order element is the reflection around the diagonal
vertical plane
that connects the right-front of the cube to the left-back
of the cube. An example of a 3-order element
is given by a (counter-clockwise) rotation of 120 degrees with axis
the line going from
the upper right corner at the front to the lower left corner at the
back of the cube.
Each of the
24 proper discrete symmetries of the cube naturally leads to
a discrete symmetry acting on the 3 vectors.
For instance, the 2-- and 3--order elements given above induce the
following discrete symmetries acting on the vectors, respectively:

\begin{eqnarray}
A^\prime &=& B\, ,\ \ \ \ B^\prime = A\, ,\ \ \ \ C^\prime = C\, ,
\nonumber\\
A^\prime &=& B\, ,\ \ \ \ B^\prime = C\, ,\ \ \ \ C^\prime = A\, .
\end{eqnarray}

To see which discrete group is realized on the 2 scalars, it is
easiest to  write the 3 vectors $\vec Q_A, \vec Q_B$ and $\vec Q_C$ as
the corners of an equilateral triangle, like in Fig.~5.
\begin{figure}[t]
\begin{center}
\mbox{\epsfig{file=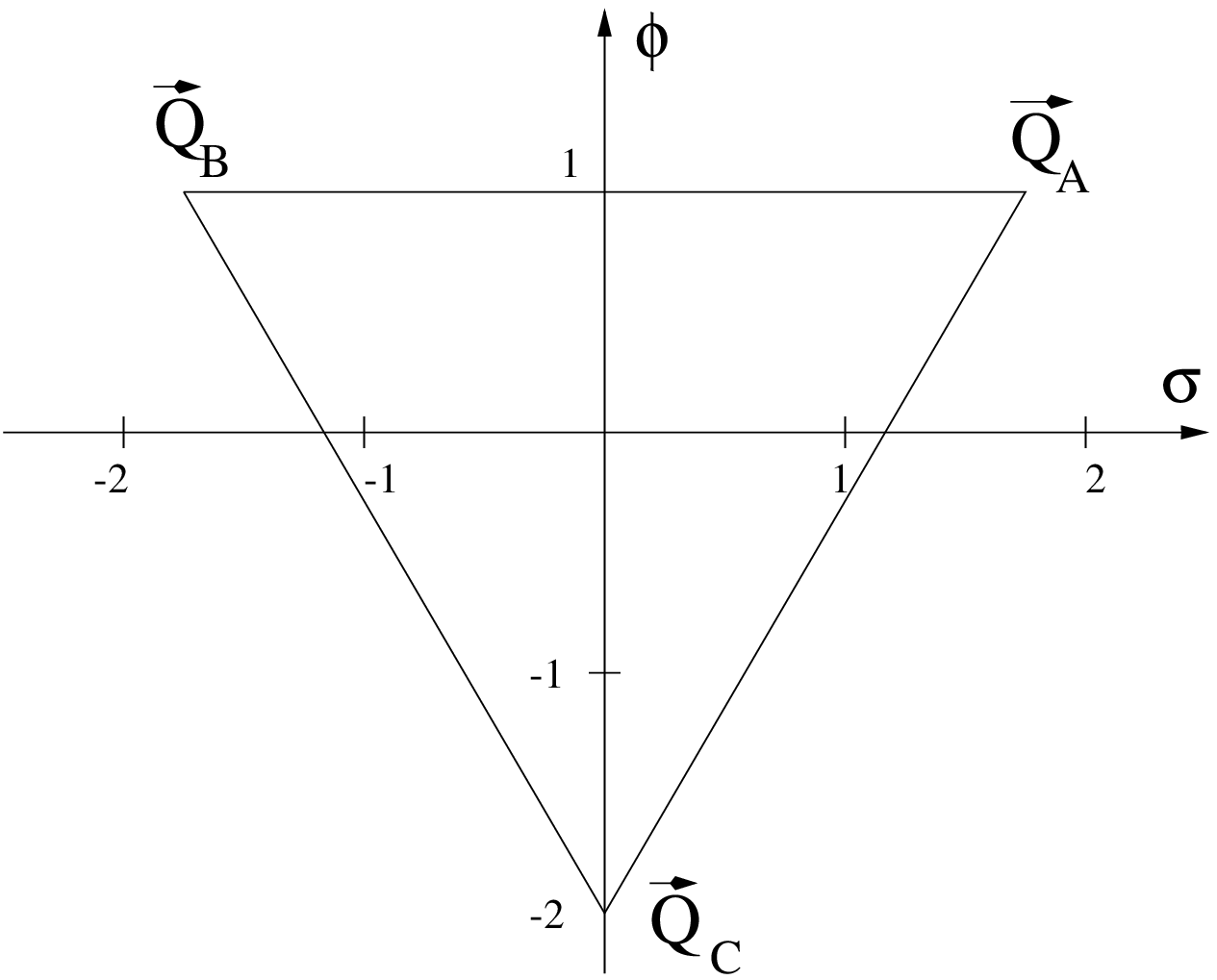, width=7cm}}
\end{center}
\noindent {\bf Fig.~5}\ \ {\it Each symmetry of the equilateral triangle
corresponds to a symmetry acting on the 2 scalars. The 3 corners of the
triangle are given by the 3 vectors $\vec Q_A, \vec Q_B$ and  $\vec Q_C$
defined in eq.~(\ref{vector}).}
\end{figure}
It was pointed out by Kaloper \cite{Ka1}
that on the scalars one can realize the 6-element dihedral group

\begin{equation}
D_3 = {\cal C}/\bigl (\Z_2\times \Z_2 \times \Z_2\bigr )\, ,
\end{equation}
i.e.~to every 4 symmetries acting on the vectors one relates a single
symmetry acting on the scalars. The action of the 6 elements of $D_3$
on the scalars is given by\footnote{
For future reference we have given each element a name. Note that
the ${\bf T}$--element corresponds to the usual $T$-duality
and that the ${\bf S}$--element corresponds to the string/string duality.
Furthermore, by ${\bf TS}$ we mean the symmetry that is obtained by
a composition of ${\bf T}$ and ${\bf S}$ as follows:

\begin{equation}
\sigma^{\prime\prime} = \tfrac{1}{2}\sigma^\prime +\tfrac{1}{2}\sqrt 3
\phi^\prime = -\tfrac{1}{2}\sigma +\tfrac{1}{2}\sqrt 3\phi\, .
\end{equation}
}

\begin{eqnarray}
\label{D3duality}
{\bf e}:\hskip 1.5truecm \sigma^\prime &=& \sigma\, ,\nonumber\\
                   \phi^\prime &=& \phi\, ,\nonumber\\
{\bf T}:\hskip 1.5truecm    \sigma^\prime &=& - \sigma\, ,\nonumber\\
                   \phi^\prime &=& \phi\, ,\nonumber\\
{\bf S}:\hskip 1.5truecm    \sigma^\prime &=& \tfrac{1}{2} \sigma +
                      \tfrac{1}{2}\sqrt 3\phi\, ,\nonumber\\
                    \phi^\prime &=& \tfrac{1}{2}\sqrt 3\sigma -
                       \tfrac{1}{2}\phi\, ,\nonumber\\
{\bf TS}:\hskip 1.5truecm   \sigma^\prime &=& -\tfrac{1}{2} \sigma +
                      \tfrac{1}{2}\sqrt 3\phi\, ,\nonumber\\
                    \phi^\prime &=& -\tfrac{1}{2}\sqrt 3\sigma -
                       \tfrac{1}{2}\phi\, ,\\
{\bf ST}:\hskip 1.5truecm   \sigma^\prime &=& -\tfrac{1}{2} \sigma -
                      \tfrac{1}{2}\sqrt 3\phi\, ,\nonumber\\
                    \phi^\prime &=& +\tfrac{1}{2}\sqrt 3\sigma -
                       \tfrac{1}{2}\phi\, ,\nonumber\\
{\bf TST}:\hskip 1.5truecm   \sigma^\prime &=& \tfrac{1}{2} \sigma -
                      \tfrac{1}{2}\sqrt 3\phi\, ,\nonumber\\
                    \phi^\prime &=& -\tfrac{1}{2}\sqrt 3\sigma -
                       \tfrac{1}{2}\phi\, .\nonumber
\end{eqnarray}

Instead of realizing the discrete symmetries on the 2 scalars
$\sigma$ and $\phi$ one may alternatively describe them by their
action on the compactification radius $R$ and the string coupling
constant $g_s$ which are related to $\sigma$ and $\phi$ via

\begin{equation}
R = e^{-2\sigma/\sqrt 3}\, ,\hskip 1.5truecm g_s =
e^{\phi - \sigma/\sqrt 3}\, .
\end{equation}
In terms of $R$ and $g_s$ the $D_3$-rules are given by\footnote{
Some of these rules have already been given in \cite{Wi1}.
Note that the ${\bf T}$--element inverts the compactification radius $R$
while the ${\bf S}$--element inverts the string coupling constant $g_s$,
as expected.}

\begin{eqnarray}
{\bf e}:\hskip 1.5truecm R^\prime &=& R\, ,\ \ g_s^\prime = g_s\, ,\nonumber\\
{\bf T}: \hskip 1.5truecm   R^\prime &=& {1\over R}\, ,\ \ g_s^\prime =
{g_s\over R}\, ,\nonumber\\
{\bf S}: \hskip 1.5truecm    R^\prime &=& {R\over g_s}\, ,\ \
g_s^\prime = {1\over g_s}\, ,\nonumber\\
{\bf TS}: \hskip 1.5truecm R^\prime &=& {1\over g_s}\, ,\ \ g_s^\prime =
{R\over
 g_s}\, ,\\
{\bf ST}:\hskip 1.5truecm                R^\prime &=& {g_s\over R}\, ,\ \
 g_s^\prime = {1\over R}\, ,\nonumber\\
{\bf TST}: \hskip 1.5truecm R^\prime &=& g_s\, ,\ \ g_s^\prime = R\, .\nonumber
\end{eqnarray}

To every element of $D_3$ corresponds 4 elements of ${\cal C}/\Z_2$ acting
on the 3 vectors. The specific transformations of the vectors are given
by\footnote{We only give 6 elements of ${\cal C}/\Z_2$. To every element
below one can associate 3 more elements  by changing (in 3 possible ways)
{\it two} signs in the given transformation rules.}

\begin{eqnarray}
\label{cubicduality}
{\bf e}: \hskip 1.5truecm A^\prime &=& A\, ,\ \ \ B^\prime = B\, ,\ \ \
C^\prime = C\, ,
\nonumber\\
{\bf T}: \hskip 1.5truecm A^\prime &=& B\, ,\ \ \ B^\prime = A\, ,\ \ \
C^\prime = C\, ,
\nonumber\\
{\bf S}: \hskip 1.5truecm A^\prime &=& A\, ,\ \ \ B^\prime = C\, ,\ \ \
C^\prime = B\, ,
\nonumber\\
{\bf TS}: \hskip 1.5truecm A^\prime &=& B\, ,\ \ \ B^\prime = C\, ,\ \ \
C^\prime = A\, ,
\\
{\bf ST}: \hskip 1.5truecm A^\prime &=& C\, ,\ \ \ B^\prime = A\, ,\ \ \
C^\prime = B\, ,
\nonumber\\
{\bf TST}: \hskip 1.5truecm A^\prime &=& C\, ,\ \ \ B^\prime = B\, ,\ \ \
C^\prime = A\, .
\nonumber
\end{eqnarray}

Finally, for the convenience of the reader we collect below some useful
properties of $D_3$. The
dihedral group $D_3$ is defined by the property that it has a 2-order
element $a$ and a 3-order element $b$ such that $ab$ is 2-order, i.e.

\begin{equation}
a^2 = b^3 = (ab)^2 = e\, ,
\end{equation}
where $e$ is the unit element. We may take $a={\bf T}$ and $b={\bf ST}$.
Note that
${\bf e}$ is 1-order, {\bf S,T,TST} are 2-order and ${\bf ST,TS}$ are
3-order. Some useful relations are

\begin{equation}
\label{D3id}
{\bf (ST)^2 = TS\, , \  \ (TS)^2=ST\, ,\ \ STS=TST\, .}
\end{equation}
The complete group multiplication table of $D_3$ is given in Table 2
below.
\bigskip
\bigskip

\begin{center}
\begin{tabular}{|c||c|c|c|c|c|c|}
\hline
&$e$&T&S&ST&TS&TST\cr
\hline\hline
$e$&$e$&T&S&ST&TS&TST\cr
\hline
T&T&$e$&TS&TST&S&ST\cr
\hline
S&S&ST&$e$&T&TST&TS\cr
\hline
ST&ST&TST&T&TS&$e$&T\cr
\hline
TS&TS&S&TST&$e$&ST&S\cr
\hline
TST&TST&TS&ST&S&T&$e$\cr
\hline
\end{tabular}
\end{center}
\bigskip

\noindent {\bf Table 2.}\ \ {\it Group multiplication table of the
6-element dihedral group $D_3$.}
\bigskip

This concludes our discussion of the discrete duality symmetries that
act in the common sector. Clearly, any of the symmetries given
above can be formulated as a 6-dimensional duality symmetry by using
the inverse of the reduction formulae given in eq.~(\ref{d=6d=5}).
We next discuss the Type II dualities. For that purpose we
will first introduce in the next section
the different theories involved.
The Type II dualities\footnote{In $D=10$
the name Type I, Type II is related to
$N=1, N=2$ supersymmetry, respectively. In $D=6$ we have $N=1,2$ and $N=4$
supersymmetry. Since the theories we will discuss have $N=2$
supersymmetry we denote their duality symmetries as Type II dualities.}
that act within and between these theories will be
discussed in section 5.
\bigskip

\noindent {\bf 3. Heterotic, Type IIA and Type IIB in Six Dimensions}
\bigskip

In this section we will describe and discuss the action and
non-compact symmetries of
(i) the 10-dimensional heterotic string compactified on $T^4$,
(ii) the 10-dimensional Type IIA string compactified on $K3$ and
(iii) the 10-dimensional Type IIB string compactified on $K3$.
The dimensional reduction of these theories from $D=6$ to $D=5$
will be discussed in the next section.
\bigskip

\noindent{\it A. Heterotic}
\bigskip

The field content of the heterotic theory is given by the usual
metric, dilaton, antisymmetric tensor system $\{\hat g, \hat \phi, \hat B\}$
plus 80 scalars and 24 Abelian gauge fields. The 80
scalars parametrize an $O(4,20)/\bigl (O(4)\times O(20)\bigr )$ coset and
are combined into
the symmetric $24 \times 24$ dimensional matrix ${\hat M}^{-1}$ satisfying
${\hat M}^{-1}L{\hat M}^{-1} = L$ where $L$ is the invariant metric on
$O(4,20)$:

\begin{equation}
\label{L420}
L = \left\{ \begin{array}{ccc}
0&\I_4&0\\
\I_4&0&0\\
0&0&-\I_{16} \end{array} \right\}\, .
\end{equation}
The heterotic action in the string-frame metric is given by:

\begin{eqnarray}
\label{h6}
I_H &=& \tfrac{1}{2}\int d^6x \sqrt {-\hat g}e^{-2\hat\phi} \biggl [-\hat R +
4|d\hat \phi|^2 - \tfrac{3}{4}|\hat H|^2\\
&&+\tfrac{1}{8}{\rm Tr}\ \bigl ( \partial_{\hat\mu} {\hat M}
\partial^{\hat\mu} {\hat M}^{-1}\bigr )
- (d \hat V)_{\hat\mu\hat\nu}^a {\hat M}_{ab}^{-1}
(d\hat V)^{\hat\mu\hat\nu b}\biggr ]\, ,\nonumber
\end{eqnarray}
where $\hat H$ is defined by

\begin{equation}
\hat H = d\hat B + {\hat V}^a d{\hat V}^b L_{ab}\, .
\end{equation}
The Chern-Simons term inside $\hat H$ is related to the following
transformation rule of $\hat B$ under the Abelian gauge transformations
with parameter $\hat \eta$:

\begin{equation}
\delta \hat B = - {\hat V}^a d {\hat \eta}^b L_{ab}\, .
\end{equation}

The heterotic action (\ref{h6}) is invariant under a non-compact
$O(4,20)$ symmetry

\begin{equation}
{\hat V}_{\hat \mu}^\prime = \Omega {\hat V}_{\hat \mu}\, ,
\hskip 1.5truecm
({\hat M}^{-1})^\prime = \Omega {\hat M}^{-1}\Omega^T\, ,
\end{equation}
where $\Omega$ is an element of $O(4,20)$.
\bigskip

\noindent{\it B. Type IIA}
\bigskip

The field content of the Type IIA theory in 6 dimensions
is identical to the heterotic
theory. The action, however, is different. Instead of a Chern-Simons
term inside $\hat H$ the action contains an additional
topological term as compared to the heterotic action. We thus
have\footnote{Although we are dealing with a different theory we
will indicate the Type IIA fields with the same symbols as the heterotic
fields. Whenever confusion could arise, we will explicitly denote
whether a field is to be considered as a heterotic or a Type IIA field.}

\begin{eqnarray}
\label{2a6}
I_{\rm IIA} &=& \tfrac{1}{2}
\int_{{\cal M}_6} d^6x \sqrt {-\hat g}e^{-2\hat\phi}
\biggl [-\hat R +  4|d\hat \phi|^2 - \tfrac{3}{4}|d\hat B|^2\\
&&+\tfrac{1}{8}{\rm Tr}\ \bigl ( \partial_{\hat\mu} {\hat M}
\partial^{\hat\mu} {\hat M}^{-1}\bigr )
-  e^{2\hat\phi}(d \hat V)_{\hat\mu\hat\nu}^a
{\hat M}_{ab}^{-1}
(d\hat V)^{\hat\mu\hat\nu b}\biggr ]\nonumber\\
&&-\tfrac{1}{8}\int_{{\cal M}_7} {d\hat B}d {\hat V}^a d{\hat V}^b
L_{ab}\, ,\nonumber
\end{eqnarray}
where ${\cal M}_7$ is a 7-manifold with boundary ${\cal M}_6$.
\bigskip

\noindent{\it C. Type IIB}
\bigskip

The field content of the Type IIB theory is given by a metric, 5
self-dual antisymmetric tensors, 21 anti--selfdual antisymmetric tensors
and 105 scalars. The 105 scalars
parametrize an $O(5,21)/\bigl (O(5) \times O(21)\bigr )$
coset and are combined into
the symmetric $26 \times 26$ dimensional matrix ${\hat {\cal M}}$
satisfying the condition
${\hat {\cal M}}^{-1} {\cal L} {\hat {\cal M}}^{-1} =
{\cal L}$ where ${\cal L}$ is the invariant metric on $O(5,21)$:

\begin{equation}
\label{m521}
{\cal L} = \left\{
\begin{array}{ccc}
0&1&0\\
1&0&0\\
0&0&L
\end{array}
\right\}\, ,
\end{equation}
and $L$ is the invariant metric on $O(4,20)$ given in (\ref{L420}).
The Type IIB theory is obtained by a reduction over $K3$ of the
10--dimensional Type IIB theory \cite{To1}. Its field equations were
constructed in \cite{Ro1}.

A complicating feature of the Type IIB theory is that, due to the self-duality
conditions on the antisymmetric tensors, there is no Lorentz-covariant
action \cite{Mar1}. However, there exists a so-called non--selfdual action that
has the property that its field equations lead to the correct
Type IIB field equations upon substituting the self-duality
constraints by hand. A similar action has been introduced for
the $D=10$ Type IIB theory \cite{Be3}. The non-self-dual action
also occurs naturally in the group-manifold approach, both in
$D=6$ \cite{Fr1} as well as $D=10$ \cite{Ca1}\footnote{In the
group manifold approach the selfduality constraints follow from
varying a non-spacetime field. Put differently, only those spacetime
background fields are allowed for a description using the group manifold
approach that satisfy the self-duality constraints.}.
An important property of the non-self-dual action is
that it leads to the
correct action for the Type IIB theory {\it after} dimensional reduction.
Therefore, the non-self-dual Type IIB action is well suited for our purposes.
The reason for this is that the self-duality conditions after the
dimensional reduction become algebraic conditions that can be substituted
back into the effective action.
We will show how this works in more detail in the next section.
We find that in the Einstein-frame the non-self-dual
Type IIB action is given by

\begin{equation}
\label{IIBaction}
I_{\rm IIB} =
\tfrac{1}{2}\int d^6 x {\sqrt {-{\hat g}}}\biggl [-\hat R +\tfrac{1}{8}
{\rm Tr}\bigl (\partial_{\hat\mu}{\hat {\cal M}}\partial^{\hat\mu}
{\hat {\cal M}}^{-1}\bigr )
+\tfrac{3}{8} (d\hat B)^i_{\hat\mu\hat\nu\hat\rho} {\hat {\cal M}}^{-1}_{ij}
(d\hat B)^{\hat\mu\hat\nu\hat\rho j}\biggr ]\, .
\end{equation}
The field--equations corresponding to this action lead to the correct
Type IIB field--equations provided we substitute by hand the following
(anti-) self--duality conditions for the
antisymmetric tensors ${\hat B}^i\ (i=1,\cdots ,26)$:

\begin{equation}
\label{asd}
(d\hat B)^i = {\cal L}^{ij} {\hat {\cal M}}^{-1}_{jk}\  {}^* (d\hat B)^k\, .
\end{equation}

In order to extract the common sector out of the
Type IIB theory, it is necessary to use a particular parametrization of
the matrix ${\hat {\cal M}}^{-1}$ in terms of the 105 scalars, thereby
identifying a particular scalar as the Type IIB dilaton $\hat\phi$.
This dilaton may then be used to define a string-frame metric ${\hat g}_S$
via ${\hat g}_S = e^{\hat\phi} {\hat g}_E$ where ${\hat g}_E$ is the
Einstein-frame metric.
We use the following parametrization:

\begin{equation}
{\hat {\cal M}}^{-1} =
\left\{\begin{array}{ccc}
-e^{-2{\hat \phi}} + {\hat \ell}^a {\hat \ell}^b {\hat { M}}^{-1}_{ab}
-{1\over 4}e^{2{\hat \phi}}{\hat\ell}^4&
{1\over 2}e^{2{\hat\phi}}{\hat\ell}^2&
{\hat\ell}^a{\hat M}^{-1}_{ab} - {1\over 2}e^{2{\hat\phi}}{\hat\ell}^2
{\hat\ell}^a L_{ab}\cr
{1\over 2}e^{2{\hat\phi}}{\hat\ell}^2&
-e^{2\hat\phi}&
e^{2{\hat\phi}}{\hat\ell}^aL_{ab}\cr
{\hat\ell}^a{\hat M}^{-1}_{ab} - {1\over 2}e^{2{\hat\phi}}{\hat\ell}^2
{\hat\ell}^a L_{ab}&
e^{2{\hat\phi}}{\hat\ell}^aL_{ab}&
{\hat M}_{ab}^{-1} - e^{2\hat\phi}{\hat\ell}^c{\hat\ell}^dL_{ac}L_{bd}
\end{array}\right\}
\end{equation}
where 80 scalars are contained in the $O(4,20)$ matrix ${\hat M}^{-1}$,
24 scalars are described by the ${\hat \ell}^a$ and where $\hat \phi$ is
identified as the Type IIB dilaton. Furthermore, we have used the definition

\begin{equation}
{\hat\ell}^2 \equiv {\hat\ell}^a{\hat\ell}^bL_{ab}\, ,\hskip 1.5truecm
(a=1,\cdots, 24)\, .
\end{equation}

The common sector is obtained by imposing the constraints:

\begin{equation}
\label{constraints}
{\hat B}_{\hat\mu\hat\nu}^i=0\, ,\ \ (i=3,\cdots ,26)\, ,\ \
{\hat \ell}^a=0\, ,\ \ \ {\hat M}^{-1}_{ab} = \delta_{ab}\, .
\end{equation}
After imposing these constraints the (anti-) self--duality conditions
(\ref{asd})
reduce to

\begin{equation}
\label{ascommon}
d{\hat B}^{(2)} = - e^{-2{\hat\phi}}\ {}^*d{\hat B}^{(1)}\, .
\end{equation}
Substituting the constraints (\ref{constraints}) and the constraint
(\ref{ascommon}) back into the Type IIB
action (\ref{IIBaction}) one obtains the standard
form of the action for the common sector in the Einstein metric as given in
(\ref{commone}). Having identified the
Type IIB dilaton it is
straighforward to convert this result to the string-frame metric as given in
(\ref{commons}). Note that although the Type IIB theory
has no Lorentz-covariant action, the common subsector does. This is due to
the fact that the corresponding (anti-) selfduality constraint (\ref{ascommon})
is off-diagonal
and hence can be used to eliminate e.g.~${\hat B}^{(2)}$ in terms of
${\hat B}^{(1)}$.
\bigskip

\noindent {\bf 4. Dimensional Reduction}
\bigskip

In this section we describe the dimensional reduction to $D=5$
of the $D=6$ theories described in the previous section.
As a main result we will show that by using particular dimensional reduction
schemes in each case the $D=6$ heterotic, Type IIA and Type IIB theories
can be mapped onto the {\it same} $D=5$ Type II theory. For reasons,
explained in the introduction, we will associate each of the
reduction formulae with an element of $D_3$. The names of
the reduction formulae (and their inverse) are
chosen such that, when restricted to the common subsector,
each reduction formula can be obtained from that of the
heterotic string (which we have chosen as the unit element) by
the action of the corresponding element of $D_3$. In the next section we
will use the reduction formulae, derived in this section,
as building blocks to construct
the different discrete duality symmetries acting in 6 dimensions.

\bigskip

\noindent{\it A. Heterotic}
\bigskip

We make the following Ansatz for the 6-dimensional fields in terms of
the 5-dimensional fields\footnote{In this section we will always use
the string-frame metric
both in $D=6$ as well as in  $D=5$.}:

\begin{equation}
\label{H5}
{\bf e:}
\left\{
\begin{array}{rcl}
{\hat g}_{\underline{xx}} &=& - e^{-4\sigma/{\sqrt 3}}\, ,\\
{\hat g}_{{\underline x}\mu} &=& -e^{-4\sigma/{\sqrt
 3}}A_\mu\, ,\\
{\hat g}_{\mu\nu} &=& g_{\mu\nu} - e^{-4\sigma/{\sqrt 3}}A_\mu
 A_\nu\, ,\\
{\hat B}_{\mu\nu} &=& B^{(C)}_{\mu\nu} + A_{[\mu}B_{\nu]}
+ \ell^a V_{[\mu}^bA_{\nu]}L_{ab}\, ,\\
{\hat B}_{\underline x \mu} &=& B_\mu -\tfrac{1}{2}\ell^a V_\mu^b
 L_{ab}\, ,\\
\hat \phi &=& \phi - \tfrac{1}{{\sqrt 3}} \sigma\, ,\\
{\hat V}_\mu^a &=& V_\mu^a +\ell^aA_\mu\, ,\\
{\hat V}_{\underline x}^a &=& \ell^a\, ,\\
{\hat M} &=& M\, .
\end{array}
\right.
\end{equation}
The Latin superscript
of the 5-dimensional antisymmetric tensor indicates to which vector this tensor
needs to be dualized in order to actually obtain the common $D=5$
Type II theory. This dualization goes via the formula

\begin{equation}
\label{duality1}
H^{\mu\nu\rho(C)} = {1\over 3\sqrt g}e^{2\phi}
\epsilon^{\mu\nu\rho\lambda\sigma}(dC)_{\lambda\sigma}\, .
\end{equation}

The inverse relations corresponding to (\ref{H5}) are given by:

\begin{equation}
\label{H5inv}
{\bf e^{-1}:}
\left\{
\begin{array}{rcl}
g_{\mu\nu} &=& {\hat g}_{\mu\nu} - {\hat g}_{{\underline x}\mu}
{\hat g}_{{\underline x}\nu}/{\hat g}_{\underline {xx}}\, ,\\
B^{(C)}_{\mu\nu} &=& {\hat B}_{\mu\nu} +{\hat g}_{{\underline
 x}[\mu}{\hat B}_{\nu]\underline x}/{\hat g}_{\underline {xx}} +
\tfrac{1}{2} {\hat V}_{\underline x}^a{\hat g}_{\underline x [\mu}{\hat
 V}^b_{\nu]}L_{ab}/{\hat g}_{\underline {xx}}\, ,\\
\phi &=& {\hat \phi} -\tfrac{1}{4}{\rm log} (-{\hat
 g}_{\underline {xx}})\, ,\\
A_\mu &=& {\hat g}_{\underline x\mu}/{\hat
 g}_{\underline {xx}}\, ,\\
B_\mu &=& {\hat B}_{\underline x\mu} + \tfrac{1}{2}{\hat
 V}_{\underline x}^a {\hat V}_\mu^b L_{ab} - \tfrac{1}{2}{\hat
 V}_{\underline x}^a {\hat V}_{\underline x}^b L_{ab}{\hat
 g}_{\underline {x}\mu}/{\hat g}_{\underline {xx}}\, ,\\
V_\mu^a &=& {\hat V}_\mu^a - {\hat V}_{\underline x}^a {\hat
 g}_{\underline x\mu}/{\hat g}_{\underline {xx}}\, ,\\
\sigma &=& -\tfrac{1}{4}\sqrt 3\ {\rm log}(-{\hat g}_{\underline {xx}})
\, ,\\
\ell^a &=& {\hat V}_{\underline x}^a\, ,\\
M &=& {\hat M}\, .
\end{array}
\right.
\end{equation}
\vskip .5truecm

The dimensionally reduced action in the (5-dimensional)
string-frame metric is given by

\begin{eqnarray}
\label{mothers}
I_{II} = &&\tfrac{1}{2} \int_{{\cal M}_5} d^5 x\sqrt
 g e^{-2\phi}\biggl\{ -R + 4 |d\phi|^2 + \tfrac{1}{8}{\rm Tr}\bigl (
\partial_\mu {\cal M}\partial^\mu {\cal M}^{-1}\bigr )\nonumber\\
&&+ e^{4\phi}|dC|^2 -
(d{\cal A})_{\mu\nu}^i {\cal M}_{ij}^{-1}(d{\cal A})^{\mu\nu j}\biggr\}\\
&& - \tfrac{1}{4}\int_{{\cal M}_6}dCd{\cal A}^i d{\cal A}^j {\cal
 L}_{ij}\, ,\nonumber
\end{eqnarray}
where ${\cal M}_6$ is a 6-manifold with boundary ${\cal M}_5$ and
${\cal L}$ is the invariant metric
on $O(5,21)$  given in (\ref{m521}).
The vectors ${\cal A}^i\ \ (i=1,\cdots ,26)$ are given by

\begin{equation}
{\cal A}_\mu^i =
\left\{
\begin{array}{ccc}
A_\mu\\
B_\mu\\
V_\mu^a
\end{array}
\right\}\, .
\end{equation}
The explicit expression of the $O(5,21)$
matrix ${\cal M}$ in terms of the 105 scalars $\sigma, \ell^a$
and the 80 scalars contained in the $O(4,20)$ matrix $M$ is given by

\begin{equation}
\begin{array}{l}
{\cal M}^{-1} =\\
\left\{ \begin{array}{ccc}
-e^{-4\sigma/\sqrt 3} + {\ell}^a {\ell}^b { { M}}^{-1}_{ab}
-{1\over 4}e^{4\sigma/\sqrt 3}{\ell}^4&
{1\over 2}e^{4\sigma/\sqrt 3}{\ell}^2&
{\ell}^a{ M}^{-1}_{ab} - {1\over 2}e^{4\sigma/\sqrt 3}{\ell}^2
{\ell}^a L_{ab}\cr
{1\over 2}e^{4\sigma/\sqrt 3}{\ell}^2&
-e^{4\sigma/\sqrt 3}&
e^{4\sigma/\sqrt 3}{\ell}^aL_{ab}\cr
{\ell}^a{ M}^{-1}_{ab} - {1\over 2}e^{4\sigma/\sqrt 3}{\ell}^2
{\ell}^a L_{ab}&
e^{4\sigma/\sqrt 3}{\ell}^aL_{ab}&
{M}_{ab}^{-1} - e^{4\sigma/\sqrt 3}{\ell}^c{\ell}^dL_{ac}L_{bd}
\end{array}\right\}
\end{array}
\end{equation}
where $\ell^2 \equiv \ell^a\ell^b L_{ab}$.
These scalars parametrize the coset
$O(5,21)/\bigl (O(5)\times O(21)\bigr )$.

The action (\ref{mothers}) defines the Type II theory in 5
dimensions\footnote{The string dynamics corresponding to this $D=5$ Type II
theory has been analyzed recently in \cite{An1,Kar1}.}.
It clearly contains the common sector given
in (\ref{common5v}). This may be seen by imposing the following constraints:

\begin{equation}
\ell^a = V_\mu^a = 0\, ,\ \ \ \ \ M_{ab}^{-1} = \delta_{ab}\, .
\end{equation}

Finally, the $D=5$ type II action in the (5-dimensional) Einstein-frame
metric takes the following form:

\begin{eqnarray}
\label{mothere}
I_{II} = &&\tfrac{1}{2} \int_{{\cal M}_5} d^5 x\sqrt
 g \biggl\{ -R - \tfrac{4}{3} |d\phi|^2 + \tfrac{1}{8}{\rm Tr}\bigl (
\partial_\mu {\cal M}\partial^\mu {\cal M}^{-1}\bigr )\nonumber\\
&&+ e^{8\phi/3}|dC|^2 - e^{-4\phi/3}
(d{\cal A})_{\mu\nu}^i {\cal M}_{ij}^{-1}(d{\cal A})^{\mu\nu j}\biggr\}\\
&& - \tfrac{1}{4}\int_{{\cal M}_6}dCd{\cal A}^i d{\cal A}^j {\cal
 L}_{ij}\, .\nonumber
\end{eqnarray}
\bigskip

\noindent{\it B. Type IIA}
\bigskip

Clearly, the dimensional reduction of the Type IIA theory leads to
a Type II theory in 5 dimensions similar to the one given above.
We also know that the
Type II theory in 5 dimensions is {\it unique}. Therefore, it should
be possible to map the $D=6$ Type IIA theory onto the {\it same} $D=5$ Type
II theory that we obtained above by dimensional reduction of the
heterotic theory. To achieve this, one must use a particular reduction
scheme of the Type IIA theory that is {\it different} from that of the
heterotic theory described above.

To be precise, we find that the dimensional reduction of the
$D=6$ Type IIA theory leads to exactly
the {\it same} $D=5$ Type II theory defined in
eq.~(\ref{mothers}) provided we use the following dimensional
reduction formulae for the Type IIA theory:

\begin{equation}
\label{IIA5}
{\bf S:}
\left\{
\begin{array}{rcl}
{\hat g}_{\underline{xx}} &=& - e^{-2\phi -2\sigma/{\sqrt 3}}\, ,\\
{\hat g}_{{\underline x}\mu} &=& -e^{-2\phi -2\sigma/{\sqrt
 3}}A_\mu\, ,\\
{\hat g}_{\mu\nu} &=& e^{-2\phi + 2\sigma/{\sqrt 3}} g_{\mu\nu} - e^{
-2\phi -2\sigma/{\sqrt 3}}A_\mu
 A_\nu\, ,\\
{\hat B}_{\mu\nu} &=& B^{(B)}_{\mu\nu} + A_{[\mu}C_{\nu]}\, ,\\
{\hat B}_{\underline x \mu} &=& C_\mu\, ,\\
\hat \phi &=& -\phi + \tfrac{1}{{\sqrt 3}} \sigma\, ,\\
{\hat V}_\mu^a &=& V_\mu^a +\ell^aA_\mu\, ,\\
{\hat V}_{\underline x}^a &=& \ell^a\, ,\\
{\hat M} &=& M\, .
\end{array}
\right.
\end{equation}
The 5-dimensional antisymmetric tensor $B_{\mu\nu}^{(B)}$ is dualised
to a vector $B_\mu$ via the relation

\begin{equation}
\label{duality2}
H^{\mu\nu\rho(B)} = {1\over 3\sqrt g} e^{2\phi +4\sigma/\sqrt 3}
\epsilon^{\mu\nu\rho\lambda\sigma}\biggl [
(dB)_{\lambda\sigma} + \ell^a(dV)_{\lambda\sigma}^b L_{ab} +
\ell^2(dA)_{\lambda\sigma}\biggr ]\, .
\end{equation}
The inverse relations corresponding to (\ref{IIA5}) are given by:

\begin{equation}
\label{IIA5inv}
{\bf S^{-1}:}
\left\{
\begin{array}{rcl}
g_{\mu\nu} &=& e^{-2\hat\phi}\biggl (
{\hat g}_{\mu\nu} - {\hat g}_{{\underline x}\mu}
{\hat g}_{{\underline x}\nu}/{\hat g}_{\underline {xx}}\biggr )\, ,\\
B^{(B)}_{\mu\nu} &=& {\hat B}_{\mu\nu} +{\hat g}_{{\underline
 x}[\mu}{\hat B}_{\nu]\underline x}/{\hat g}_{\underline {xx}} \, ,\\
\phi &=& -\tfrac{1}{2}{\hat \phi} -\tfrac{1}{4}{\rm log} (-{\hat
 g}_{\underline {xx}})\, ,\\
A_\mu &=& {\hat g}_{\underline x\mu}/{\hat
 g}_{\underline {xx}}\, ,\\
C_\mu &=& {\hat B}_{\underline x\mu}\, ,\\
V_\mu^a &=& {\hat V}_\mu^a - {\hat V}_{\underline x}^a {\hat
 g}_{\underline x\mu}/{\hat g}_{\underline {xx}}\, ,\\
\sigma &=& \tfrac{{\sqrt 3}}{2}\hat\phi
-\tfrac{{\sqrt 3}}{4} {\rm log}(-{\hat g}_{\underline {xx}})
\, ,\\
\ell^a &=& {\hat V}_{\underline x}^a\, ,\\
M &=& {\hat M}\, .
\end{array}
\right.
\end{equation}

\noindent{\it C. Type IIB}
\bigskip

The above discussion for the Type IIA theory also applies to the
Type IIB theory. We find that the dimensional reduction of the Type IIB
 theory leads to
 the {\it same} $D=5$ Type II theory (\ref{mothers})
as the dimensional reduction
 of the heterotic and Type IIA theory provided we use the following
dimensional reduction formulae for the Type IIB fields:

\begin{equation}
\label{IIB}
{\bf ST:}
\left\{
\begin{array}{rcl}
{\hat g}_{\underline{xx}} &=& - e^{2\phi +2\sigma/{\sqrt 3}}\, ,\\
{\hat g}_{{\underline x}\mu} &=& -e^{2\phi +2\sigma/{\sqrt
 3}}C_\mu\, ,\\
{\hat g}_{\mu\nu} &=& e^{-2\phi + 2\sigma/{\sqrt 3}} g_{\mu\nu} - e^{
2\phi +2\sigma/{\sqrt 3}}C_\mu
 C_\nu\, ,\\
{\hat B}_{\mu\nu}^i &=& B^{({\cal A})i}_{\mu\nu} + C_{[\mu}
{\cal A}^i_{\nu]}\, ,\\
{\hat B}^i_{\underline x \mu} &=& {\cal A}^i_\mu\, ,\\
\hat \phi &=& \tfrac{2}{{\sqrt 3}} \sigma\, ,\\
{\hat \ell}^a &=& \ell^a\, ,\\
{\hat M} &=& M\, .
\end{array}
\right.
\end{equation}

Note that due to the (anti-)
self--duality relations (\ref{asd}) both ${\hat B}_{\mu\nu}^i$
as well as ${\hat B}_{{\underline x}\mu}^i$ get related to the
5-dimensional vector fields ${\cal A}_\mu^i$. This goes as follows.
In a first stage the dimensional reduction of ${\hat B}_{\mu\nu}^i$
leads to a 5-dimensional antisymmetric tensor $B_{\mu\nu}^{({\cal B})i}$
that may be dualized to a 5-dimensional vector ${\cal B}_\mu^i$.
To perform this dualization one adds an extra term to the kinetic term
of $B_{\mu\nu}^{({\cal B})i}$ containing ${\cal B}_\mu^i$ as
a Lagrange multiplier for the Bianchi identity of $B_{\mu\nu}^{({\cal B})i}$.
The two terms together are given by

\begin{equation}
\label{2terms}
\tfrac{3}{16}{\sqrt g}e^{2\phi}H_{\mu\nu\rho}^i{\cal M}^{-1}_{ij}
H^{j\mu\nu\rho} +\tfrac{1}{8}\epsilon^{\mu\nu\rho\lambda\sigma}
{\cal B}_\mu^i\biggl (\partial_\nu H_{\rho\lambda\sigma}^j
-2\partial_\nu C_\rho \partial_\lambda {\cal A}_\sigma^j\biggr )
{\cal L}_{ij}\, ,
\end{equation}
where $H^i$ is now considered as an ${\it independent}$ field.
The field equation of $H_{\mu\nu\rho}^i$ leads to the identity

\begin{equation}
\label{feH}
H^{k\rho\sigma\lambda} = -{1\over 3\sqrt g}e^{-2\phi}
\epsilon^{\mu\nu\rho\sigma\lambda}{\cal M}^{ki}(d {\cal B})_{\mu\nu}^j
{\cal L}_{ij}\, .
\end{equation}
At the same time, the dimensional reduction of the $D=6$
(anti-) selfduality constraint (\ref{asd}) leads to exactly the
same identity in $D=5$ except that ${\cal B}_\mu^i$ is replaced
by ${\cal A}_\mu^i$:

\begin{equation}
H^{k\rho\sigma\lambda} = -{1\over 3 \sqrt g}e^{-2\phi}
\epsilon^{\mu\nu\rho\sigma\lambda}{\cal M}^{ki}(d {\cal A})_{\mu\nu}^j
{\cal L}_{ij}\, .
\end{equation}
Combining these two results leads to the conclusion that the original
$D=6$ constraint (\ref{asd}) has become a simple algebraic relation
in $D=5$:

\begin{equation}
{\cal B}_\mu^i = {\cal A}_\mu^i\, .
\end{equation}
Substituting the field equation of $H_{\mu\nu\rho}^i$, eq.~(\ref{feH}),
with the above identification understood,
back into (\ref{2terms}) then leads to the desired result. Note that
the last term in (\ref{2terms}) leads to the topological term which
is present in the 5-dimensional Type II action (\ref{mothers}).

Finally, the inverse relations are given by

\begin{equation}
\label{IIB5inv}
{\bf (ST)^{-1}:}
\left\{
\begin{array}{rcl}
g_{\mu\nu} &=& -e^{-2\hat\phi}{\hat g}_{\underline {xx}}
\biggl (
{\hat g}_{\mu\nu} - {\hat g}_{{\underline x}\mu}
{\hat g}_{{\underline x}\nu}/{\hat g}_{\underline {xx}}\biggr )
\, ,\\
B^{({\cal A})i}_{\mu\nu} &=& {\hat B}^i_{\mu\nu} +{\hat g}_{{\underline
 x}[\mu}{\hat B}^i_{\nu]\underline x}/{\hat g}_{\underline {xx}}
\, ,\\
\phi &=& -\tfrac{1}{2}{\hat \phi} +\tfrac{1}{2}{\rm log} (-{\hat
 g}_{\underline {xx}})\, ,\\
{\cal A}^i_\mu &=& {\hat B}^i_{\underline x\mu}\, ,\\
\sigma &=& \tfrac{1}{2}\sqrt 3\hat\phi\, ,\\
\ell^a &=& {\hat \ell}^a\, ,\\
M &=& {\hat M}\, .
\end{array}
\right.
\end{equation}

\bigskip

\noindent {\bf 5. Type II Dualities}
\bigskip

In this section we will show how the reduction formulae constructed in the
previous section may be used in a systematic way to construct discrete
duality symmetries in 6 dimensions that act within and between the
heterotic, Type IIA and Type IIB theories.

We first note that to each reduction formula given in the previous section
one can associate 3 further reduction formulae. This is due to the
fact that although the $D=5$ Type II theory given in ({\ref{mothers})
is not invariant under the full 24-element
proper cubic group it is still
invariant under a 4-element $\Z_2\times \Z_2$ subgroup, i.e.

\begin{equation}
\label{z2z2}
{\cal C}/\Z_2 \rightarrow \Z_2 \times \Z_2\, .
\end{equation}
This formula is the 5-dimensional analogue of the $D=9$ formula
given in (\ref{d=9sb}). In particular the $T$-duality symmetry
remains unbroken\footnote{Note that this is {\it different} from the situation
in 9 dimensions where the $T$--duality symmetry {\it is} broken.}.
Its explicit form in $D=5$ is given by  the following
particular $\Z_2$ subgroup of the non-compact $O(5,21)$ symmetry
with parameter $\Omega$ given by:

\begin{equation}
\Omega = {\cal L}\, ,
\end{equation}
where ${\cal L}$ is the flat metric given in eq.~(\ref{m521}).
In terms of components the $D=5$ $T$-duality rules
in the string-frame metric read:

\begin{eqnarray}
g_{\mu\nu}^\prime &=& g_{\mu\nu}\, ,\nonumber\\
\phi^\prime &=& \phi\, ,\nonumber\\
e^{-4\sigma^\prime/\sqrt 3} &=& e^{-4\sigma/\sqrt 3}\biggl (
e^{-8\sigma/\sqrt 3} - e^{-4\sigma/\sqrt 3}\ell^a\ell^b M_{ab}^{-1}
+ \tfrac{1}{4}\ell^4\biggr )^{-1}\, ,\nonumber\\
\ell^{\prime a} &=& {e^{-4\sigma/\sqrt 3}\ell^c M_{cd}^{-1}L^{da}
-\tfrac{1}{2}\ell^2\ell^a\over
e^{-8\sigma/\sqrt 3} - e^{-4\sigma/\sqrt 3}\ell^c\ell^d M_{cd}^{-1}
+\tfrac{1}{4}\ell^4}\, ,\nonumber\\
M^\prime &=& M^{-1}\, ,\\
A_\mu^\prime &=& B_\mu\, ,\nonumber\\
B_\mu^\prime &=& A_\mu\, ,\nonumber\\
C_\mu^\prime &=& C_\mu\, ,\nonumber\\
V_\mu^{\prime} &=& L V_\mu\, .\nonumber
\end{eqnarray}
Note that, when restricted to the common sector these rules reduce to the
standard ones given in section 2. We also observe that, for the special case
that $M_{ab} = \delta_{ab}$, the complicated duality rules of $\sigma$ and
$\ell^a$ factorize such that the final result can be written in terms
of an effective metric as was discussed in \cite{Be2}. This effective
metric has a natural origin in a sigma model approach \cite{Hu2}.
Apparently this factorization does not occur for nontrivial values of
the matrix $M$.

The second $\Z_2$-factor in (\ref{z2z2}) corresponds to another
$\Z_2$ subgroup of $O(5,21)$ with parameter $\Omega$ given by

\begin{equation}
\Omega = - \I\, .
\end{equation}
Its action in components is given by

\begin{equation}
{\cal A}_\mu^{\prime i} = - {\cal A}_\mu^i\, ,
\end{equation}
while all other fields remain invariant. To simplify the
discussion below we will from now on only concentrate on those
transformations that act non-trivially on the scalars.
In that
case we are only dealing with the dihedral group $D_3$ which gets broken to:

\begin{equation}
D_3 \rightarrow \Z_2\, ,
\end{equation}
where the $\Z_2$-factor is given by the $T$-duality transformation above.
The presence of this $T$-duality symmetry in 5 dimensions means that
to each reduction formula we can associate a so-called $T$-dual version.
Its explicit form is obtained by replacing in the original reduction
formula each 5-dimensional fields by its $T$-dual expression.
The $T$-dual reduction formula so obtained should lead to the
same answer in 5
dimensions. This is guaranteed by the fact that the 5-dimensional
action is invariant under $T$-duality. We will indicate the
$T$--dual versions of the reduction formulae constructed in the previous
section as follows:

\begin{eqnarray}
\label{Tred}
{\bf e} & \rightarrow & {\bf T}\, ,\nonumber\\
{\bf S} &\rightarrow & {\bf TS}\, ,\\
{\bf ST} &\rightarrow & {\bf TST}\, .\nonumber
\end{eqnarray}

We thus obtain 6 different reduction formulae which correspond to the
3 down-pointing arrows in Figure 6. Similarly, there are 6 inverse
reduction, or so-called oxidation, formulae which go opposite
the vertical arrows in Figure 6. These oxidation formulae will be indicated
by the inverse group elements. The claim is now that, using these
6 reduction and oxidation formulae only,
one is able to construct in a simple way all
the discrete dualities that act within and between the heterotic, Type
IIA and Type IIB theories that are indicated in Figure 6.
\begin{figure}[t]
\begin{center}
\mbox{\epsfig{file=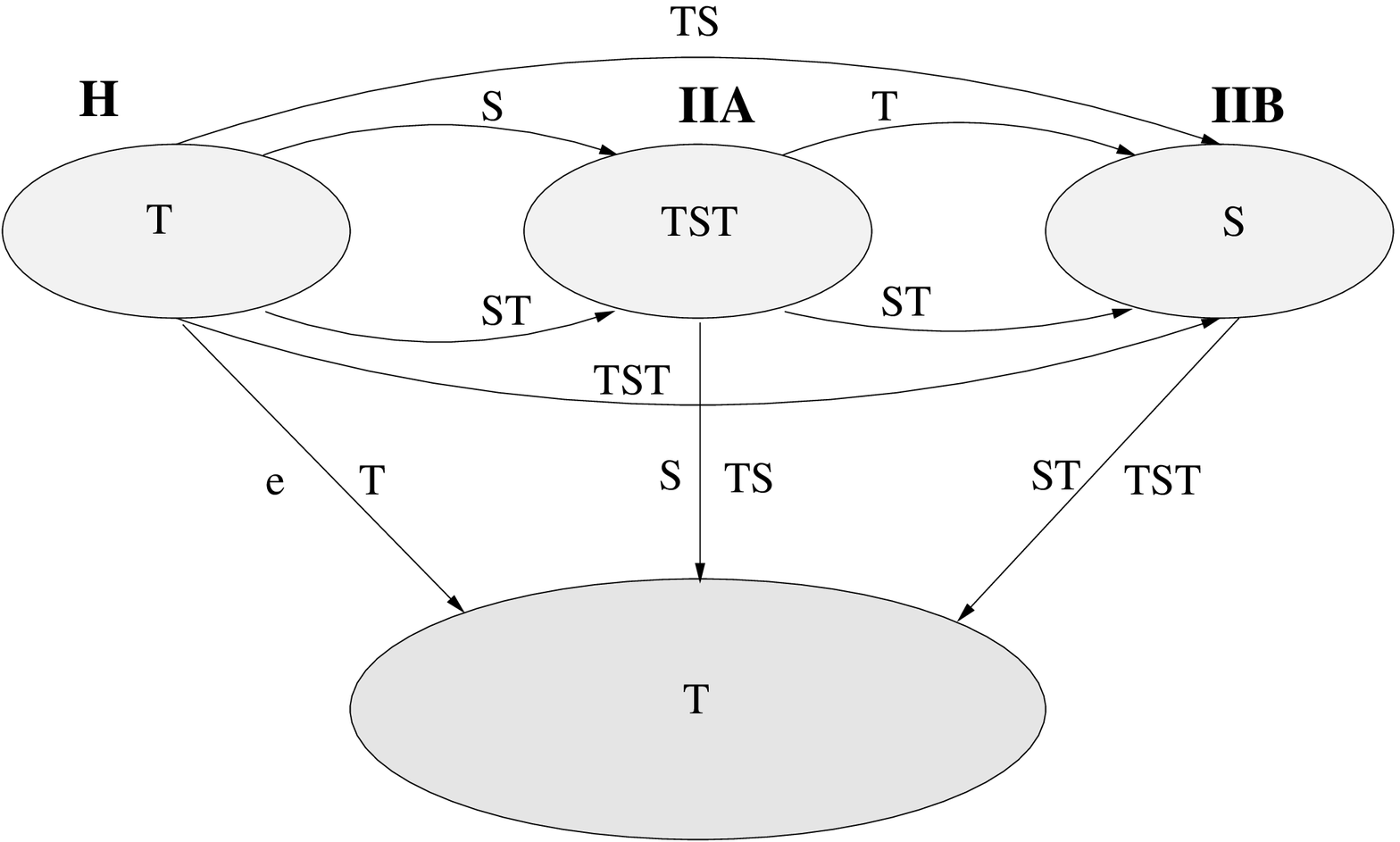, width=10cm}}
\end{center}
\noindent {\bf Fig.~6}\ \ {\it The 3 down-pointing arrows indicate
the 6 possible ways to map the 3 $D=6$ theories (heterotic, Type IIA,
Type IIB) onto the {\it same} $D=5$ Type II theory. Each reduction
formula is indicated by a (boldface) element of $D_3$. As explained in the
text these 6 reduction formula and their inverses
may be used to construct the explicit form
of all the discrete $D=6$ dualities that are indicated in the figure.}
\end{figure}
Each discrete
duality symmetry has been given a name such that, when restricted to
the common subsector, the duality becomes the corresponding $D_3$
duality symmetry that acts in the common subsector.

To show how the dualities of Figure 6 may be constructed starting from the
different reduction and oxidation formulae it is instructive to first
give a few examples.

\begin{enumerate}

\item The $T$-duality that acts within the heterotic theory is obtained
    by first reducing the heterotic theory using the ${\bf e}$ reduction
formulae given in (\ref{H5}) and next the oxidation formula
${\bf T^{-1}}$ defined in (\ref{Tred}), i.e.

\begin{equation}
T (H\rightarrow H) = {\bf T^{-1}} \times {\bf e} = {\bf T}\, .
\end{equation}

\item The $S$-duality that maps the heterotic onto the Type IIA
theory is obtained by first reducing the heterotic theory with ${\bf e}$ and
next oxidizing the $D=5$ theory with ${\bf S^{-1}}$. As Figure 6 shows
there are 3 other possibilities, one of them gives the same answer
while the other two are related to the $ST$ map indicated in figure 6:

\begin{eqnarray}
S(H\rightarrow IIA) &=&  {\bf S^{-1}} \times {\bf e} = {\bf S^{-1}} =
{\bf S}\, ,\nonumber\\
S(H\rightarrow IIA) &=& {\bf (TS)^{-1}} \times {\bf T} = {\bf ST}
\times {\bf T} = {\bf S} \, ,\\
(ST)(H\rightarrow IIA) &=& {\bf S^{-1}} \times {\bf T} = {\bf S}
\times {\bf T} = {\bf ST}\, ,\nonumber\\
(ST)(H\rightarrow IIA) &=& {\bf (TS)^{-1}}\times {\bf e} = {\bf ST}
\times {\bf e} = {\bf ST}\, .\nonumber
\end{eqnarray}
Note that we have used here the group multiplication table of $D_3$
given in Table 2. The reason that we are allowed to use the group
multiplication table of $D_3$ is that we have set up our notation for
the 6 reduction and oxidation formulae in such a way that, when
restricted to the
common subsector, these reduction and oxidation
formulae actually become specific $D_3$ symmetries in 5 dimensions.

\item
The $S$-duality that acts within the
IIB theory is obtained by first reducing the
IIB theory with ${\bf ST}$ and then oxidizing with ${\bf (TST)^{-1}}$.
The other way round gives the same answer:

\begin{eqnarray}
S(IIB\rightarrow IIB) &=& {\bf (TST)^{-1}}\times {\bf ST} =
{\bf TST}\times {\bf ST} \nonumber\\
&=& {\bf STS} \times {\bf ST} =  {\bf S}\, ,\\
S(IIB\rightarrow IIB) &=& {\bf (ST)^{-1}}\times {\bf TST} =
{\bf TS}\times {\bf TST} \nonumber\\
&=& {\bf TS} \times {\bf STS} =  {\bf S}\, ,\nonumber
\end{eqnarray}
where we have used some of the $D_3$ identities given in eq.~(\ref{D3id}).

\item
We deduce from figure 6 that there is not only a $T$-duality
that acts within the heterotic theory but also a $T$-duality that
maps the IIA theory onto the IIB theory. It may be
obtained in the following 2 ways from the reduction/oxidation formulae:

\begin{eqnarray}
T(IIA\rightarrow IIB) &=& {\bf (ST)^{-1}} \times {\bf S} =
{\bf TS}\times {\bf S} = {\bf T}\, ,\nonumber\\
T(IIA\rightarrow IIB) &=& {\bf (TST)^{-1}}\times {\bf TS}
= {\bf TST}\times {\bf TS} = {\bf T}\, .
\end{eqnarray}

\item
Finally, we observe that $ST$ is a 3-order element of $D_3$.
This means that starting with the heterotic theory and applying the
$ST$-duality 3 times we should get back the heterotic theory.
In the diagram of figure 6 this is seen as follows: The 1st $ST$
duality brings us to the IIA theory, the 2nd one brings us from the IIA
to the IIB theory. Finally, to perform the last $ST$ duality we
observe that $ST = (TS)^{-1}$, i.e.~this duality brings us back
from the Type IIB theory to the heterotic theory via the opposite
direction of the oriented arrow at the top of the diagram.
\end{enumerate}

The above examples should explain the main idea of how the reduction
and oxidation formulae of the previous section are used to construct the
different discrete duality symmetries in 6 dimensions. As a further
illustration of our method we will now give the explicit expression of
3 of such $D=6$ dualities.

\begin{description}

\item[(A)] Clearly, the $S$-duality map from the heterotic onto the Type
IIA theory should reproduce the known $D=6$ string/string duality rule
\cite{Du1}. It indeed does and we find that the $S$ duality is given by
(using the string-frame metric):

\begin{eqnarray}
 {\hat G}_{\hat\mu\hat\nu} &=& e^{-2\hat\phi}
{\hat g}_{\hat\mu\hat\nu}\, ,\nonumber\\
{\hat \Phi} &=& - \hat\phi\, ,\\
{\hat H}_{\hat\mu\hat\nu\hat\rho} &=& e^{-2\hat\phi}\ {}^*
{\hat h}_{\hat\mu\hat\nu\hat\rho}\, ,\nonumber
\end{eqnarray}
where the other fields are invariant and
where the capital fields are Type IIA and the smallscript fields heterotic.
To derive this string/string duality rule one must also use the
two dualization formulae (\ref{duality1}) and (\ref{duality2}). Note
that one may only derive a string/string duality rule for $\hat H$
and not $\hat B$. This is of course related to the fact that from the
6-dimensional point of view the string/string duality is a symmetry
of the equations of motion only.

\item[(B)] The only other $D=6$ duality that is purely $S$, and hence can
be written in a 6-dimensional covariant way\footnote{The reason that
the $S$-duality rules can be written in a 6-dimensional
covariant way, i.e.~in terms of the $\hat \mu$--indices,
is that, in contrast to the $T$-duality,
their presence does {\it not} require
the existence of a special isometry direction.}, is the one that acts within
the Type IIB theory. We find that this is given by a
particular $O(5,21)$ transformation with parameter $\Omega$ given by

\begin{equation}
\Omega = {\cal L}\, ,
\end{equation}
where ${\cal L}$ is the flat $O(5,21)$ metric given in eq.~(\ref{m521}).
In components its action on the antisymmetric tensors and scalars is
given by

\begin{eqnarray}
{\hat H}_{\hat\mu\hat\nu\hat\rho}^{\prime (1)} &=&
{\hat H}_{\hat\mu\hat\nu\hat\rho}^{(2)}\, ,\nonumber\\
{\hat H}_{\hat\mu\hat\nu\hat\rho}^{\prime (2)} &=&
{\hat H}_{\hat\mu\hat\nu\hat\rho}^{(1)}\, ,\nonumber\\
{\hat H}_{\hat\mu\hat\nu\hat\rho}^{\prime a} &=&
\bigl (L{\hat H}_{\hat\mu\hat\nu\hat\rho}\bigr )^a\, ,\\
e^{-2{\hat\phi}^\prime} &=& e^{-2\hat\phi}\biggl (e^{-4\hat\phi}
-e^{-2\hat\phi^\prime}{\hat\ell}^a{\hat\ell}^b{\hat M}_{ab}^{-1} +
\tfrac{1}{4}\ell^4\biggr )^{-1}\, ,\nonumber\\
{\hat\ell}^{\prime a} &=& {e^{-2\hat\phi}{\hat \ell}^c
{\hat M}_{cd}^{-1}L^{da} -\tfrac{1}{2}{\hat\ell}^2{\hat\ell}^a\over
e^{-4\hat\phi} - e^{-2\hat\phi}{\hat\ell}^a{\hat\ell}^b{\hat M}_{ab}^{-1}
+\tfrac{1}{4}{\hat\ell}^4}\, ,\nonumber\\
{\hat M} &=& {\hat M}^{-1}\, .\nonumber
\end{eqnarray}
Note that, when restricted to the common subsector, this duality
transformation indeed reduces to the standard $S$-duality rule
given in section 2.

\item[(C)] We finally give an example of a discrete duality that
involves a $T$-duality and hence cannot be written in a
6-dimensional covariant. The example
we consider concerns the $T$-duality map from the
Type IIA onto the Type IIB theory. Following our method described
above we find the following expression for this duality transformation:

\begin{eqnarray}
{\hat \Phi} &=& \hat \phi -\tfrac{1}{2}{\rm log}
(-{\hat g}_{\underline {xx}})\, ,\nonumber\\
{\hat G}_{\underline {xx}} &=& 1/{\hat g}_{\underline {xx}}\, ,
\nonumber\\
{\hat G}_{\underline x \mu} &=& {\hat b}_{\underline x\mu}/
{\hat g}_{\underline {xx}}\, ,\nonumber\\
{\hat G}_{\mu\nu} &=& {\hat g}_{\mu\nu} - \biggl (
{\hat g}_{\underline x\mu}{\hat g}_{\underline x\nu} - {\hat b}_{
\underline x\mu}{\hat b}_{\underline x\nu}\biggr )/{\hat g}_{\underline {xx}}
\, ,\\
{\hat B}_{\underline x\mu}^{(1)} &=& {\hat g}_{\underline x\mu}/
{\hat g}_{\underline {xx}} \, ,\nonumber\\
{\hat B}_{\mu\nu}^{(1)} &=& {\hat b}_{\mu\nu} -  \biggl (
{\hat g}_{\underline x\mu} {\hat b}_{\underline x\nu} -
{\hat g}_{\underline x\nu} {\hat b}_{\underline x\mu}\biggr )
/ {\hat g}_{\underline {xx}} \, ,\nonumber\\
{\hat B}_{\underline x\mu}^a &=& {\hat v}_\mu^a - {\hat v}_{\underline x}^a
{\hat g}_{\underline x\mu}/{\hat g}_{\underline {xx}} \, ,\nonumber\\
{\hat\ell}^a &=& {\hat v}_{\underline x}^a\, ,\nonumber\\
{\hat M}_{ab} &=& {\hat m}_{ab}\, ,\nonumber
\end{eqnarray}
where the capital  fields are IIB and the smallscript fields are
IIA fields, respectively. Note that the duality transformations of
${\hat B}^{(2)}_{\hat\mu\hat\nu}$ and
${\hat B}_{\mu\nu}^a$ are not given. Their transformation rules
follow from the ones given above via the the self-duality conditions
(\ref{asd}).
\end{description}

\bigskip

\noindent {\bf 6. The 6-dimensional chiral null model}
\renewcommand{\arraystretch}{1.6}
\bigskip

To give the reader a better understanding of how the
discrete duality symmetries constructed in this paper
relate different backgrounds to each other we discuss in this section
the chiral null model as a special example. We will present
here only some aspects related to the common sector. A more
detailed discussion, including the Type II sector, will be
given elsewhere \cite{Be4}. Our starting point is the reduction of the $D=6$
common sector to $D=5$ which defines the {\bf e} element.
Next, we oxidize back to 6 dimensions
in two different ways, related to the {\bf $\bf S^{-1}$} and ${\bf
(TST)^{-1}}$ element indicated in figure 6. Starting with the
ciral null model, these two group elements
lead to two other 6-dimensional solutions. We thus end up with 3 different
6-dimensional solutions which are all dual to each other. We will
describe these 3 solutions below.
\bigskip

\noindent
{\it A. The {\bf e} element}
\medskip

\noindent
The chiral null model is a string background that allows one conserved
chiral current on the world sheet. As a D=10 heterotic string background
it has unbroken supersymmetries and is exact in the $\alpha'$ expansion
\cite{Ho1}\footnote{The issue of unbroken supersymmetry and
$\alpha^\prime$--corrections for special
cases of the $D=10$ chiral null model has been studied in \cite{Be5}.}.
The dimensional reduction yield many known black hole and
Taub-Nut geometries \cite{Kal1,Beh}.  Since we are interested here only in
the common sector we assume a trivial reduction from D=10 to D=6 and
thus obtain the following expression for the $D=6$ chiral null model
(in the string--frame metric):
\begin{equation}  \label{chir}
\begin{array}{l}
d{\hat{s}}^2 = 2 F(x) \, du \left[ dv - \frac{1}{2} K(x) \, du +
  \omega_I dx^I \right] - dx^I dx^I  \\
\hat{B} = - 2 F(x) \,du \wedge \left[ dv + \omega_I dx^I \right]
\qquad , \qquad e^{2 \hat{\phi}} = F(x)\, ,
\end{array}
\end{equation}
where $u,v$ are standard light--cone coordinates,
$F^{-1}$ and $K$ are harmonic functions and $\omega_I$ fulfills
the Maxwell equation with respect to the transversal coordinates $x^I$
($I,J = 1,..4$):
\begin{equation}
\partial^2 F^{-1} = \partial^2 K = \partial_I \partial_{[I} \omega_{J]}
 =0 \ .
\end{equation}
In the case of $\omega_I =0$ this model is an interpolation between
the gravitational wave background ($F=1$) and fundamental string
($K=1$)\footnote{The standard parameterization of the fundamental string
is given by $K=0$. However, this would lead to a singularity since
we have to invert $K$ below. For this reason we make use of
the possibility to give $v$ a linear shift. We thus obtain $K=1$
as a fundamental string. Physically this means that we give the string a
linear momentum (see \cite{Ho1}).}.

Using the reduction formulae given in eq.~(\ref{d=6d=5}) we reduce
this solution to 5 dimensions.  If we assume that the internal direction
is $u$ (note: $\hat{g}_{\underline{xx}} = {\hat g}_{uu} < 0$,
i.e.\ space-like)
and $v$ is the time we find for the fields in the Einstein frame
\begin{equation}    \label{5dsol}
\begin{array}{ccc}
ds^2 = (\frac{F}{K})^{\frac{2}{3}} \left(dt + \omega_I dx^I\right)^2 -
(\frac{K}{F})^{\frac{1}{3}} dx_I dx_I & , &
H_{\mu\nu\rho} = 2 F K \, A_{[\mu} \partial_{\nu} A_{\rho]} \\
e^{-\frac{4}{\sqrt{3}} \sigma} =  FK & ,
& e^{-4 \phi} = \frac{K}{F} \\
A_{\mu} = - K^{-1} ( \, 1\, , \, \omega_I\,) &,& B_{\mu} = - F (\, 1\, , \,
  \omega_I)\, .
\end{array}
\end{equation}
These fields define a solution of the 5-dimensional type II theory.
The 6-dimensional solution related to the {\bf e} element is given by
the original $D=6$ ciral null model given in (\ref{chir}).
To construct the other dual solutions we have to oxidize
the $D=5$ solution in different ways. Before we can do that we have to
dualize the torsion which defines the third vector $C_{\mu}$. The result
(in Einstein-frame) is
\begin{equation}   \label{dualh}
\begin{array}{ll}
F_{\sigma\tau}(C) = \partial_{\sigma} C_{\tau} - \partial_{\tau} C_{\sigma}
 &= \frac{1}{2} \frac{1}{\sqrt{g}} \,
e^{-\frac{8}{3} \phi} \,
  g_{\sigma\alpha}\,  g_{\tau\beta} \, \epsilon^{\alpha\beta\mu\nu\rho}
 H_{\mu\nu\rho} \\
 &= \frac{1}{2} \delta_{\sigma I} \, \delta _{\tau J}\,
\epsilon^{IJKL} \partial_{K} \omega_{L},
\end{array}
\end{equation}
where we have used eq.~(\ref{duality1}).
To make the situation more transparent we restrict ourselves to the
static limit (diagonal metric): $\omega_I = 0$, which has the
consequence that
\begin{equation}
C_{\mu}=0\ .
\end{equation}

\bigskip
\pagebreak

\noindent
{\it B. The $ {\bf S^{-1}}$ element}
\medskip

\noindent
This element was related to the field redefinition (see (\ref{D3duality})
and (\ref{cubicduality}))
\begin{equation}
\sigma' = {1\over 2} (\sigma + \sqrt 3\phi)  \qquad , \qquad
\phi^\prime = {1\over 2}(\sqrt 3\sigma - \phi) \qquad \mbox{and} \qquad
B \leftrightarrow C\ .
\end{equation}
We have to insert for $\sigma$ and $\phi$ the functions given in
(\ref{5dsol}) and take $\sigma'$ for the new compactification radius and
$\phi'$ for the new dilaton. Furthermore, $B$ was the KK field coming
from the antisymmetric tensor and the interchange of the two gauge
fields means that we have no KK gauge field coming from the torsion but
we have a non-vanishing $D=5$ torsion. The new $D=6$ solution is then
given by (using the string-frame metric)
\begin{equation} \label{IIa}
\begin{array}{l}
d{\hat{s}}^2 = 2 du\, \left( dv - \frac{1}{2} K \, du \right) - F^{-1} \,
 dx_I dx_I \qquad , \qquad e^{-2 \hat{\phi}} = F \\
\hat{H}_{IJK} = \frac{1}{6} \epsilon_{IJKL} \partial^L F^{-1}\, ,
\qquad \hat{H}_{IJu} = \hat{H}_{IJv} = \hat{H}_{Iuv} = 0\, .
\end{array}
\end{equation}
This solution has been discussed already in \cite{cv/ts} and the
generalization to stationary metrics ( $\omega \neq 0$,
i.e.~non-vanishing angular momentum in $D=5$) can be found in
\cite{be/do}.  It can be interpreted as an interpolating solution
between a wave background ($F=1$) and the solitonic string ($K=1$)
solution \cite{Du1,Se1,Ha1}.

\bigskip

\noindent
{\it C. The $ {\bf (TST)^{-1}}$ element}
\medskip
\newline
\noindent
For this element we have the field redefinition (see (\ref{D3duality})
and (\ref{cubicduality}))
\begin{equation}
\sigma' =  {1\over 2} (\sigma - \sqrt 3\phi)  \qquad , \qquad
\phi^\prime = -{1\over 2}(\sqrt 3\sigma + \phi) \qquad \mbox{and} \qquad
A \leftrightarrow C\ .
\end{equation}
Since we are considering $C=0$ only, the interchange of the vectors
means here, that we have no KK gauge field coming from the metric (no
off-diagonal term). We find for the 6-dimensional solution (in the
string-frame metric)
\begin{equation}  \label{IIb}
\begin{array}{l}
d{\hat{s}}^2 = F \left( dv^2 - du^2 \right) - K \,
 dx_I dx_I \qquad , \qquad e^{2 \hat{\phi}} = F K\, , \\
\hat{H}_{IJK} = - \frac{1}{6} \epsilon_{IJKL}
\partial^L K\ , \quad  \hat{H}_{Iuv} = \partial_I F\, ,
\qquad \hat{H}_{IJu} = \hat{H}_{IJv} = 0\, .
\end{array}
\end{equation}
This string type solution was discussed before in \cite{du/fe}.
In analogy to the previous cases we can now interpret this solution as
an interpolating solution between the fundamental string ($F=1$)
and solitonic string ($K=1$). Note that in relation to the
{\bf e} element (the original solution (\ref{chir})) the dilaton
and the compactification radius (${\hat g}_{uu}$) have
interchange their role.

\medskip

Finally, one can ask what about the other cases: {\bf T}, {\bf TS} or
{\bf ST}? From the previous sections we
know that these cases correspond to the
5-dimensional solution that is $T$-dual to our 5-dimensional starting
point (\ref{5dsol}). From (\ref{D3duality}) and (\ref{cubicduality})
we find the replacements
\begin{equation}
\sigma' = -\sigma  \qquad , \qquad
\phi^\prime = \phi \qquad \mbox{and} \qquad
A \leftrightarrow B \ ,
\end{equation}
which simply means that we have to interchange both harmonic functions
\begin{equation}
K \leftrightarrow F^{-1} \ .
\end{equation}
This, however, does not change the structure of the 6-dimensional
solutions. Thus, these elements are related to internal symmetries of every
6-dimensional solution whereas the other elements ({\bf S} and {\bf TST})
correspond to solution generating transformations.

\medskip

Inserting now harmonic functions for $K$ and $F^{-1}$ we first find
that all three solutions are asymptotically free and that near the
singularity the {\bf e} solution is in the weak coupling region, the
{\bf S} solution is strongly coupled whereas for the {\bf TST}
solution the string coupling constant remains finite. In a forthcoming
paper \cite{Be4} we will discuss the charge and mass spectrum of these
solutions.  Also, we will add 6-dimensional gauge fields and
generalize them to truly heterotic (\ref{chir}), type IIA (\ref{IIa})
and type IIB (\ref{IIb}) solutions with non trivial RR fields.

\renewcommand{\arraystretch}{1.0}

\bigskip

\noindent {\bf Acknowledgements}
\vspace{.5truecm}

We would like to thank M.~Bianchi, H.J.~Boonstra, G.~Papadopoulos,
A.~Sagnotti, P.~Townsend and A.~Tseytlin for discussions.
In particular we would
like to thank Tomas Ort\'\i n who was involved in the initial discussions
leading to the present work.
One of us (K.B.) would
like to thank Groningen University for hospitality.
The work of K.B.~is supported by a grant of the DFG.
The work of E.B.~has been made possible by a fellowship of the Royal
Netherlands Academy of Arts and Sciences (KNAW).
The work of B.J.~ was performed as part of the research programme of the
``Stichting voor Fundamenteel Onderzoek der Materie'' (FOM).

\end{document}